\documentclass[11pt,authoryear,review,3p,times]{elsarticle}
\usepackage[colorlinks=true,citecolor=blue,linkcolor=blue]{hyperref}
\usepackage[english]{babel}
\usepackage[utf8]{inputenc}
\usepackage{float}
\usepackage{lscape}
\usepackage[T1]{fontenc}
\usepackage{graphicx,array,bm,color,natbib,amssymb}
\usepackage{latexsym,verbatim}
\usepackage[dvips]{psfrag}
\usepackage{subfigure}
\bibpunct{(}{)}{;}{a}{\hspace{-0.1em},}{,} 
\usepackage{amsmath}
\usepackage{mathrsfs}
\usepackage{latexsym,verbatim}
\usepackage{theorem}
\usepackage{enumerate}

\newcommand{\dev}{$r_i^{dev}$}
\newcommand{\pea}{$r_i^{pea}$}

\newcommand{\wil}{$r_i^{wil}$}
\newcommand{\qu}{$r_i^{qu}$}

\newcommand{\aasc}{$r_i^{ans}$}

\newcommand{\aqu}{$r_i^{*qu}$}
\newdefinition{definition}{Definition} 
\newproof{proof}{Proof} 
\newproof{prop}{Property} 
\graphicspath{{figures/}{../figures/}} 
\renewcommand{\baselinestretch}{1.25} 
\makeatletter
\def\@author#1{\g@addto@macro\elsauthors{\normalsize%
    \def\baselinestretch{1}%
    \upshape\authorsep#1\unskip\textsuperscript{%
      \ifx\@fnmark\@empty\else\unskip\sep\@fnmark\let\sep=,\fi
      \ifx\@corref\@empty\else\unskip\sep\@corref\let\sep=,\fi
      }%
    \def\authorsep{\unskip,\space}%
    \global\let\@fnmark\@empty
    \global\let\@corref\@empty  %% Added
    \global\let\sep\@empty}%
    \@eadauthor={#1}
}
\makeatother

\makeatletter
\def\ps@pprintTitle{%
   \let\@oddhead\@empty
   \let\@evenhead\@empty
   \let\@oddfoot\@empty
   \let\@evenfoot\@oddfoot
}
\makeatother

\begin{document}
		
		\begin{frontmatter}
		\title{{ \LARGE Adjusted quantile residual for generalized linear models }}

		\author{\large Juliana Scudilio \fnref{label1,label3}\corref{aaa}}
%    \author{\large Juliana Scudilio \fnref{label1,label3}$^\star$}	
		\author{\large Gustavo H. A. Pereira\fnref{label3}}	
	
		\address[label1]{Maths Science Institute and Computing, University of S\~ao Paulo,  S\~ao Carlos, SP, Brazil.}
		\address[label3]{ Department of Statistics, Federal University of S\~ao Carlos,  S\~ao Carlos, SP, Brazil.}

		\cortext[aaa]{e-mail: juliana-scudilio@uol.com.br\\ }

\begin{abstract}
Generalized linear models are widely used in many areas of knowledge. As in other classes of regression models, it is desirable to perform diagnostic analysis
in generalized linear models using residuals that are approximately standard normally distributed.
Diagnostic analysis in this class of models are usually performed using the standardized Pearson residual or the standardized deviance residual. The former has skewed distribution and the latter has negative mean, specially when the variance of the response variable is high. In this work, we introduce the adjusted quantile residual for generalized linear models. Using Monte Carlo simulation techniques and two applications, we compare this residual with the standardized Pearson residual, the standardized deviance residual and two other residuals. Overall, the results suggest that the adjusted quantile residual is a better tool for diagnostic analysis in generalized linear models.
\end{abstract}

\end{frontmatter}

\section{Introduction}

Generalized linear models (GLMs) were introduced by \cite{nelder72} and are widely used to study the relationship between a response
variable and a set of predictor variables. It assumes that the distribution of the response variable given the set of predictors belongs to the exponential family and can be used to model many types of variables, such as binary variables, count variables and continuous skewed variables. The routines to fit GLMs are implemented in many statistical softwares and these models are described in details in \cite{mccullagh} and \cite{hardin2007generalized}.

In regression models, it is important to check model adequacy and to identify outliers and influential observations. To reach these goals, it is interesting to use residuals whose distribution is well approximated by the standard normal distribution. When the residuals do not have this property, it is not unusual that they are also non-identically distributed (see, for example, Table 1 of \cite{anholeto2014}). In these cases, it is hard to interpret residuals plots and well-fitted models can be discarded (see Application 3 of \cite{pereira2017quantile}).

Many residuals are used to perform diagnostic analysis in GLMs, such as the deviance residual \citep{davison}, the Pearson residual \citep{mccullagh}, the Williams residual \citep{williams} and the Anscombe residual \citep{pierce1986residuals}. Usually they are standardized so that their variance become close to one.
However, the distributions of these residuals are not approximated standard normal when the variance of the response variable is high (see Section \ref{simu}).
Some works introduced residuals for GLMs that are approximated standard normal distributed \citep{cordeiro2004pearson,urbano2012wald}, but they are complex and are not implemented in the statistical softwares. 

The quantile residual \citep{dunn} is simple and general, but is not often used in GLMs. It is usually used to perform diagnostic analysis in complex regression models, such as the generalized additive models for location, scale and shape \citep{rigby2005}. The quantile residual is asymptotically standard normally distributed if the parameters are consistently estimated \citep{dunn}, but its distribution is unknown in small sample sizes. \cite{feng2017randomized} compared the quantile residual with the deviance residual and the Pearson residual in generalized linear models. Using simulation studies, the authors concluded that the distribution of the quantile residual is better approximated by the standard normal distribution than that of the others residuals. In addition, \cite{feng2017randomized} showed that the quantile residual is the best for detecting lack of fit. However, the authors considered sample size equal to $1000$ and they did not note that the variance of the quantile residual is far from one in small sample sizes (see Section \ref{qu}).

In this work, we introduce the adjusted quantile residual for GLMs. Using Monte Carlo simulation studies and two applications, we compare
the behavior of the adjusted quantile residual and other residuals usually used in GLMs.

This paper is organized as follows. Section \ref{method} presents the models used in this work, introduces the adjusted quantile residual and defines other residuals considered in this paper. In Section \ref{simu}, Monte Carlo simulation studies are performed to compare the properties of the defined residuals in several scenarios. In the following section, we study the ability of the residuals to
identify model misspecification and outliers. Finally, in Section \ref{conc}, we present the conclusions.

 \section{Methodology}
 \label{method}
 \subsection{Generalized linear models}
 \label{mod}
  
Let $y_1, y_2, \ldots, y_n$ be independent random variables, where each $y_i$ has a density functions in the linear exponential family defined as
  \begin{equation}
  \pi(y;\theta_i,\phi)= \exp\left[\phi \left\{ y\theta_i-b(\theta_i)+c(y)\right\}+d(y;\phi)\right],
  \label{glm}
  \end{equation}
  
  where $b(\cdot)$ and $c(\cdot)$ are know functions and $\phi_i$ is a precision parameter. It can be proved that $E(y_i)=\mu_i = b'(\theta_i)$ and $Var(y_i)=\phi^{-1}V(\mu_i)$, where $V_i=V(\mu_i)=d\mu_i/d\theta_i$ is the variance function.
  
  Generalized linear models (GLMs) are defined by the family of distributions specified in equation \ref{glm} and the systematic component
  \begin{equation*}
  g(\mu_i) = \eta_i = x_i^\top\boldsymbol{\beta},
  \end{equation*}
  where $x_{i}=(x_{i1},x_{i2},\ldots,x_{ik})^\top$ is a vector of known covariates ($k < n$), $\boldsymbol{\beta}=(\beta_1,\beta_2,\ldots,\beta_k)^\top$ is a vector of unknown parameters ($\boldsymbol{\beta} \in \mathbb{R}^k$) and $g(.)$ is a strictly monotonic and twice differentiable link function.

 This paper focus on generalized linear models with continuous and asymmetric response variables.
 We considered only continuous variable as response variable, because in the discrete case the residual distribution cannot be continuous without randomization.
We also use asymmetric distributions because in the normal case a residual with good properties is already known.
 
 The gamma regression models and the inverse Gaussian regression models are the most important regression models in the class of GLMs with asymmetric and continuous response variable.
They are considered in this paper and we define these models using the parameterization used by \cite{stasinopoulos08}.
 
The gamma distribution with parameters $\mu_i$ and $\sigma$ has a density function defined as 

\begin{eqnarray*}
\label{gama}
f_1(y;\mu_i,\sigma)=\frac{exp \left(-\frac{ y}{\sigma^2\mu_i}\right)}{\Gamma(\frac{1}{\sigma^2})} \frac{y^{(\frac{1}{\sigma^2}-1)}}{\left(\mu_i\sigma^2\right)^{\frac{1}{\sigma^2}}},
\end{eqnarray*}
 \noindent where $y>0$, $\mu_i>0$  and $\sigma>0$.
  In this parameterization of the gamma distribution, $\mbox{E}(y_i)=\mu_i$ and $\mbox{Var}(y_i)=\sigma^2\mu_i^2$.

The inverse Gaussian distribution with parameters $\mu_i$ and $\sigma$ is defined as 
 \begin{equation*}
 f_2(y;\mu_i,\sigma)=\frac{1}{\sqrt{(2\pi y^3 \sigma^2)}}exp\left\{-\frac{(y-\mu_i)^2}{2\mu_i^2\sigma^2y}\right\},
 \label{invga}
 \end{equation*}
 \noindent where $y>0$, $\mu_i>0$  and $\sigma>0$. In this parameterization of the inverse Gaussian, $\mbox{E}(y_i)=\mu_i$ and $\mbox{Var}(y_i)=\sigma^2\mu_i^3$. 
 
 The parameters of the GLMs can be estimated by maximum likelihood using numerical optimization algorithms and hypothesis tests can be conducted using Wald, score or likelihood ratio statistic.
The \textit{glm} function and the \textit{gamlss} package, which are available for the R statistical software, can be used for fitting gamma and inverse Gaussian regression models.

\subsection{Standardized quantile residual}
\label{qu}
The quantile residual can be used in a wide class of regression models. 
When the response variable is discrete, it uses uniform distributed random variables and it is called randomized quantile residual.
For the subclass of GLMs considered in this paper, it is defined as
 
 \begin{equation}
 r_i^{qu}=\Phi^{-1}\{F(y_i;\hat{\mu}_i,\hat{\sigma})\},
 \label{quanti}
 \end{equation}  
 where $\Phi(.)$ is the cumulative distribution function of the standard normal distribution, $F(.)$ is the cumulative distribution function of the gamma distribution or inverse Gaussian distribution and  $\hat{\mu}_i$ and $\hat{\sigma}$ are the maximum likelihood estimates of the parameters $\mu_i$ and $\sigma$, respectively.

 According to \cite{dunn}, \qu\, is asymptotically standard normally distributed, but it is important to study its properties when sample size is not large.
We used Monte Carlo simulation studies to study the properties of the quantile residual in small sample sizes. 
Gamma and inverse Gaussian regression models were considered and we assumed the model $log(\mu_i)=\beta_0+\beta_{1}x_{i1}+\beta_{2}x_{i2}$.  The covariate values were generated as independent draws from the standard uniform distribution and remained constant throughout the simulations.
We considered \,$\beta_{0}=3$, $\beta_1=2$ and $\beta_2=1$,  which resulted in $\mu \in (20{.}085 ; 403{.}429)$ and used $\sigma=0{.}1$ for gamma regression model and $\sigma=0{.}02$ for inverse Gaussian regression model.

The left side of Tables \ref{tab-quant} and \ref{tab-quant1} present the sample mean, variance, skewness and kurtosis coefficients for \qu\ for gamma and inverse Gaussian regression models, respectively. 
The tables also contain the value of the Anderson-Darling (AD) statistic \citep{anderson}, used to test whether each residual is standard normally distributed.
The value of the Anderson-Darling statistic is used as a closeness measure between each residual distribution and standard normal distribution. Results are based on $5000$ Monte Carlo replications. Note that for both models the variance of \qu\, is far from 1.

 % Table generated by Excel2LaTeX from sheet 'Planilha1'
 \begin{table}[h!]
 	\centering
 	\tabcolsep=0.07cm    
 	\caption{Comparison between the quantile residual and the adjusted quantile residual - gamma distribution.}
 	\scalebox{0.95}{
 		\begin{tabular}{ccccccccccccc}
 		\hline
 		&       & \multicolumn{5}{c}{\qu}               &       & \multicolumn{5}{c}{\aqu} \\
 		\cline{3-13}i     & $\mu$   & Mean  & Variance & Skewness & Kurtosis & Ad    &       & Mean  & Variance & Skewness & Kurtosis & AD Statistics \\
 		\hline
 		1     & 67.742 & -0.009 & 0.848 & -0.050 & -0.398 & 5.711 &       & -0.009 & 1.031 & -0.050 & -0.398 & 3.408 \\
 		2     & 100.306 & 0.022 & 0.839 & -0.029 & -0.462 & 7.699 &       & 0.024 & 0.984 & -0.029 & -0.462 & 4.191 \\
 		3     & 215.703 & -0.010 & 0.825 & -0.011 & -0.454 & 7.726 &       & -0.011 & 1.022 & -0.011 & -0.454 & 3.924 \\
 		4     & 115.585 & 0.003 & 0.875 & -0.006 & -0.444 & 4.013 &       & 0.003 & 0.985 & -0.006 & -0.444 & 2.781 \\
 		5     & 131.633 & 0.003 & 0.879 & -0.089 & -0.420 & 4.291 &       & 0.004 & 0.967 & -0.089 & -0.420 & 2.733 \\
 		6     & 200.681 & 0.022 & 0.802 & -0.088 & -0.335 & 13.260 &       & 0.024 & 0.978 & -0.088 & -0.335 & 3.543 \\
 		7     & 33.844 & 0.016 & 0.697 & -0.104 & -0.393 & 33.215 &       & 0.019 & 1.002 & -0.104 & -0.393 & 4.099 \\
 		8     & 80.061 & 0.010 & 0.834 & -0.032 & -0.371 & 7.544 &       & 0.011 & 0.993 & -0.032 & -0.371 & 1.759 \\
 		9     & 48.232 & 0.015 & 0.651 & -0.058 & -0.433 & 45.315 &       & 0.018 & 0.990 & -0.058 & -0.433 & 3.506 \\
 		10    & 52.030 & 0.007 & 0.723 & -0.041 & -0.374 & 25.817 &       & 0.009 & 0.985 & -0.041 & -0.374 & 2.006 \\
 		11    & 48.729 & -0.003 & 0.853 & -0.061 & -0.423 & 5.350 &       & -0.004 & 1.020 & -0.061 & -0.423 & 3.670 \\
 		12    & 114.403 & 0.010 & 0.638 & -0.012 & -0.433 & 49.026 &       & 0.013 & 1.032 & -0.012 & -0.433 & 4.688 \\
 		13    & 194.271 & -0.009 & 0.848 & -0.018 & -0.403 & 5.933 &       & -0.010 & 1.025 & -0.018 & -0.403 & 3.815 \\
 		14    & 103.731 & 0.001 & 0.927 & -0.022 & -0.427 & 1.572 &       & 0.001 & 0.998 & -0.022 & -0.427 & 2.320 \\
 		15    & 208.921 & 0.018 & 0.794 & -0.046 & -0.470 & 12.496 &       & 0.021 & 1.043 & -0.046 & -0.470 & 7.130 \\
 		\hline
 		& Mean  & 0.006 & 0.802 & -0.045 & -0.416 & 15.265 &       & 0.007 & 1.004 & -0.045 & -0.416 & 3.572 \\
 		& SD    & 0.011 & 0.086 & 0.031 & 0.037 & 15.522 &       & 0.012 & 0.023 & 0.031 & 0.037 & 1.294 \\
 		\hline	
 		\end{tabular}%
 	}
 	\label{tab-quant}%
 \end{table}%

In GLMs, Pearson and deviance residuals are standardized so that their variance becomes close to one.
Our proposal is	to use the same standardization term used for these residual in the quantile residual, ie, divide the residual \qu \, by $\sqrt{(1-\hat{h}_{ii})}$, where  $\hat{h}_{ii}$ is the $i$-th diagonal element of matrix  $\hat{\boldmath{H}}=\hat{W}^{1/2}X(X^T\hat{W}X)^{-1}X^T\hat{W}^{1/2}$ and $\hat{W}=diag\{\hat{w_1},....,\hat{w_n}\}$ is the diagonal matrix of weights.
In particular, for the gamma distribution and for the inverse Gaussian distribution, $w_i=({d\mu_i}/{d\eta_i})^2/\mu_i^2$ and $w_i=({d\mu_i}/{d\eta_i})^2/\mu_i^3$, respectively. 
 Then, the adjusted quantile residual is defined as
\begin{equation}
r_i^{*qu} = \frac{r_i^{qu}}{\sqrt{(1-\hat{h}_{ii})}}.
\end{equation}
We used the term adjusted instead of standardized because \cite{klar2012specification} introduced a different residual called standardized quantile
residual to derive a goodness-of-fit test for generalized linear models. We did not consider the standardized quantile
residual in this work because it assumes that the variance of the quantile residual is constant across observations. Tables \ref{tab-quant} and \ref{tab-quant1} suggest that this assumption is not reasonable.

% Table generated by Excel2LaTeX from sheet 'Planilha2'
\begin{table}[h!]
	\centering
	\tabcolsep=0.07cm    
	\caption{Comparison between the quantile residual and the adjusted quantile residual - inverse Gaussian distribution.}
	\scalebox{0.95}{
	\begin{tabular}{ccccccccccccc}
		\hline
		&       & \multicolumn{5}{c}{\qu}               &       & \multicolumn{5}{c}{\aqu} \\
		\cline{3-13}i     & $\mu$   & Mean  & Variance & Skewness & Kurtosis & AD    &       & Mean  & Variance & Skewness & Kurtosis & AD \\
		\hline
		1     & 67.742 & 0.007 & 0.851 & -0.060 & -0.424 & 5.708 &       & 0.008 & 1.034 & -0.060 & -0.424 & 5.000 \\
		2     & 100.306 & 0.007 & 0.864 & -0.123 & -0.307 & 5.644 &       & 0.007 & 1.003 & -0.123 & -0.307 & 2.700 \\
		3     & 215.703 & 0.039 & 0.932 & -0.171 & -0.320 & 9.226 &       & 0.042 & 1.072 & -0.171 & -0.321 & 15.513 \\
		4     & 115.585 & 0.019 & 0.935 & -0.093 & -0.369 & 3.078 &       & 0.020 & 1.053 & -0.093 & -0.369 & 6.654 \\
		5     & 131.633 & 0.034 & 0.913 & -0.111 & -0.426 & 7.442 &       & 0.036 & 1.014 & -0.111 & -0.426 & 9.541 \\
		6     & 200.681 & 0.020 & 0.901 & -0.171 & -0.241 & 5.628 &       & 0.021 & 1.035 & -0.172 & -0.240 & 6.316 \\
		7     & 33.844 & 0.054 & 0.528 & -0.165 & -0.390 & 109.272 &       & 0.074 & 1.012 & -0.166 & -0.395 & 22.802 \\
		8     & 80.061 & 0.013 & 0.870 & -0.077 & -0.379 & 5.361 &       & 0.014 & 1.028 & -0.077 & -0.379 & 5.174 \\
		9     & 48.232 & 0.007 & 0.554 & -0.131 & -0.373 & 87.104 &       & 0.010 & 0.996 & -0.130 & -0.377 & 3.868 \\
		10    & 52.030 & -0.001 & 0.684 & -0.080 & -0.372 & 35.364 &       & -0.001 & 1.001 & -0.080 & -0.372 & 2.527 \\
		11    & 48.729 & 0.012 & 0.854 & -0.095 & -0.361 & 6.530 &       & 0.013 & 1.020 & -0.095 & -0.361 & 4.495 \\
		12    & 114.403 & 0.013 & 0.725 & -0.150 & -0.333 & 26.833 &       & 0.016 & 1.028 & -0.149 & -0.341 & 6.055 \\
		13    & 194.271 & -0.003 & 0.892 & -0.150 & -0.309 & 3.629 &       & -0.003 & 1.024 & -0.150 & -0.309 & 3.474 \\
		14    & 103.731 & 0.018 & 0.953 & -0.066 & -0.412 & 2.651 &       & 0.019 & 1.038 & -0.066 & -0.412 & 5.574 \\
		15    & 208.921 & 0.001 & 0.881 & -0.157 & -0.350 & 4.847 &       & 0.001 & 1.036 & -0.156 & -0.351 & 5.324 \\
		\hline
		& Mean  & 0.016 & 0.822 & -0.120 & -0.358 & 21.221 &       & 0.018 & 1.026 & -0.120 & -0.359 & 7.001 \\
		& SD & 0.016 & 0.136 & 0.040 & 0.050 & 32.846 &       & 0.020 & 0.020 & 0.040 & 0.050 & 5.407 \\
		\hline
	\end{tabular}%
}
	\label{tab-quant1}%
\end{table}%

The results of the adjusted quantile residual are shown on the right side of Tables \ref{tab-quant} and \ref{tab-quant1}. 
Note that dividing \qu\,\, by $\sqrt{(1-\hat{h}_{ii})}$ results in a residual with mean, skewness and kurtosis similar to \qu, a variance close to $1$ and smaller AD than the \qu.
 In other words, the distribution of \aqu\, is closer to the standard normal distribution than that of the \qu.
We did the same analysis in several scenarios, which were omitted in this paper because they resulted on similar conclusions.

\subsection{Other residuals in GLMs}
\label{others}
The other residuals considered in this paper are the deviance residual, the Pearson residual, the Anscombe residual and the William residual. 
%Among these residuals, the most common residual in Gamma and inverse Gaussian regression models is the deviance residual \citep{davison}.  

The standardized deviance residual is defined as
	\begin{equation}
	r_i^{dev}=\frac{d(y_i;\hat{\mu}_i)}{\hat{\sigma}\sqrt{1-\hat{h}_{ii}}},
	\end{equation}
where $d(y_i;\hat{\mu}_i)=\left\{-log(\frac{y_i}{\hat{\mu_i}})+\frac{(y_i-\hat{\mu}_i)^2}{y_i\hat{\mu}_i^2}\right\}$ for the gamma distribution and 
$d(y_i;\hat{\mu}_i)=\left\{\frac{(y_i-\hat{\mu}_i)}{y_i\hat{\mu}_i^2}\right\}$ for the inverse Gaussian distribution.

  The standardized Pearson residual is given by 
 
 	\begin{equation}
 	r_i^{pea}=\frac{(y_i-\hat{\mu}_i)}{\hat{\sigma}\sqrt{\hat{\mu}_i^2(1-\hat{h}_{ii})}}.
 	\label{pea}
 	\end{equation}

The William residual is also used in GLMs and is defined as
	\begin{equation}
	r_{i}^{wil}=sign\{y_i-\hat{\mu}_i\}[(1-\hat{h}_{ii})(r_i^{dev})^2+\hat{h}_{ii}(r_i^{pea})^2]^{1/2}.
	\end{equation}

The last residual used in this paper is the standardized Anscombe residual given by
\begin{equation}
r_i^{ans}=\frac{\psi(y_i)-\psi(\hat{\mu}_i)}{\hat{\sigma}\hat{V}^{1/2}(\hat{\mu})\psi'(\hat{\mu}_i)\sqrt{(1-\hat{h}_{ii})}},
\end{equation}
where $\psi(\mu_i)=3\mu_i^{1/3}$ and $V(\mu_i)=\mu^2$ for the gamma distribution and $\psi(\mu_i)=log(\mu_i)$ and $V(\mu_i)=\mu^3$ for the inverse Gaussian distribution.
 
\section{Simulation studies}\label{simu}
In this section, we use Monte Carlo simulation studies to compare the properties of the residuals in the GLMs. 
Here, we use the gamma and the inverse Gaussian regression models.
We considered seven scenarios for each distribution to study the residuals' properties.
In the first six scenarios, we assumed the model as
\begin{equation*}
	log(\mu_i)=\beta_{0}+\beta_{1}x_{i1}+\beta_{2}x_{i2}.
\end{equation*}

In the first four scenarios and in the last scenario for gamma and inverse Gaussian regression models, the covariates values were generated as independent draws from the standard uniform distribution. 
For the others scenarios, $x_{i1}$ values were generated from the normal distribution, ($x_{i1}\sim$ NO $(0.5,0.25^2)$), $x_{i2}$ values were generated from the inverse Gaussian distribution in scenarios V-a and V-b ($x_{i2}\sim $ IG $ (0.4,2)$) and from the gamma distribution in scenarios VI-a and VI-b ($x_{i2}\sim $ GA $ (0.4,1)$). 
The covariates values remained constant throughout the simulations. 

In order to check if the results change considerably when other link function is used, in the last scenario we used the canonical link function. 
In Scenario VII-a, we used the gamma regression model with the inverse link function, in which
\begin{eqnarray*}
\frac{1}{\mu_i} = \beta_{0}+\beta_{1}x_{i1}+\beta_{2}x_{i2}.
\end{eqnarray*}

In the last scenario (Scenario VII-b), we used the inverse Gaussian regression model with the link function $1/\mu^2$, in which
\begin{eqnarray*}
	\frac{1}{\mu_i^2} = \beta_{0}+\beta_{1}x_{i1}+\beta_{2}x_{i2}.
\end{eqnarray*}

Tables \ref{tab:1} and \ref{tab:2} present the description of the scenarios for gamma and inverse Gaussian regression models.
In the first scenario (Scenario I-a e I-b), we considered $\beta_{0}=3$, $\beta_1=2$ and $\beta_2=1$,  which resulted in $\mu \in (20{.}085 ; 403{.}429)$ and used $\sigma=0{.}1$ for gamma regression model and $\sigma=0{.}02$ for inverse Gaussian regression model.
In the second and third scenarios, we used the same $\beta_j$, $j=0,1,2$, of the Scenario I, but with different $\sigma$ values. In scenarios II-a and II-b we decreased the variance and in scenarios III-a and III-b we increased the variance. 
In the following scenarios (scenarios IV-a and IV-b), we changed the values of the coefficients to result in mean response values close to zero.
In scenarios V and VI, we changed the generating distribution of $x_{i1}$ and $x_{i2}$.
In the last scenarios (scenarios VII-a and VII-b), we changed the link function.
For all scenarios, two different sample sizes were considered: $n=15$ and $n=50$. 
All results are based on $5000$ Monte Carlo replications. 
All simulations were performed using the software R.
Results for \qu\, were omitted from the tables, because the simulation studies showed that their distributions are worse approximated by standard normal distribution than that of \aqu\, in all scenarios. 

% Table generated by Excel2LaTeX from sheet 'Planilha1'
\begin{table}[htbp]
	\centering
	\scriptsize
	\caption{Description of the scenarios for gamma regression model.}
	\begin{tabular}{ccrrrr}
		\hline
		Scenario & \multicolumn{1}{c}{Link function} & \multicolumn{1}{c}{Coefficients} & \multicolumn{1}{c}{Covariates} &  \multicolumn{1}{c}{$\mu$} & \multicolumn{1}{c}{$\sigma$} \\
		\hline
		I-a   & $log(\mu_i)$ & $\beta_0 = 3$, $\beta_1 = 2$, $\beta_2 = 1$ & $x_{i1}\sim U(0,1); x_{i2}\sim U(0,1)$& $(20{.}085 ; 403{.}429)$ & $0{.}10$ \\
		II-a  & $log(\mu_i)$& $\beta_0 = 3$, $\beta_1 = 2$, $\beta_2 = 1$ & $x_{i1}\sim U(0,1); x_{i2}\sim U(0,1)$& $(20{.}085 ; 403{.}429)$ & $0{.}05$ \\
		III-a & $log(\mu_i)$& $\beta_0 = 3$, $\beta_1 = 2$, $\beta_2 = 1$ & $x_{i1}\sim U(0,1); x_{i2}\sim U(0,1)$& $(20{.}085 ; 403{.}429)$ & $0{.}50$ \\
		IV-a & $log(\mu_i)$ & $\beta_0 = -3$, $\beta_1 = 1{.}5$, $\beta_2 = 1$ & $x_{i1}\sim U(0,1); x_{i2}\sim U(0,1)$& $(0{.}049; 0{.}606)$ & $0{.}10$ \\
		V-a  & $log(\mu_i)$ & $\beta_0 = 3$, $\beta_1 = 2$, $\beta_2 = 1$ & $x_{i1}\sim NO(0.5,0.25^2); x_{i2}\sim IG(0.4,2)$& $(20{.}085 ; 403{.}429)$ & $0{.}10$ \\
		VI-a & $log(\mu_i)$ & $\beta_0 = 3$, $\beta_1 = 2$, $\beta_2 = 1$ & $x_{i1}\sim NO(0.5,0.25^2); x_{i2}\sim GA(0.4,1)$& $(7{.}400 ; 1892{.}000)$ & $0{.}10$ \\
		VII-a & $\frac{1}{\mu_i}$ & $\beta_0 =0{.}0025$ $\beta_1 = 0{.}04$, $\beta_2 = 0{.}01$ & $x_{i1}\sim U(0,1); x_{i2}\sim U(0,1)$& $(19{.}04 ; 4000)$ & $0{.}10$ \\
		\hline
	\end{tabular}%
	\label{tab:1}%
\end{table}%

\begin{table}[htbp]
	\vspace{-1.5em}
	\scriptsize
	\centering
	\caption{Description of the scenarios for inverse Gaussian regression model.}
	\begin{tabular}{ccrrrr}
		\hline
		Scenario & \multicolumn{1}{c}{Link function} & \multicolumn{1}{c}{Coefficients} & \multicolumn{1}{c}{Covariates} &  \multicolumn{1}{c}{$\mu$} & \multicolumn{1}{c}{$\sigma$} \\
		\hline
		I-b   & $log(\mu_i)$& $\beta_0 = 3$, $\beta_1 = 2$, $\beta_2 = 1$ &  $x_{i1}\sim U(0,1); x_{i2}\sim U(0,1)$& $(20{.}085 ; 403{.}429)$ & $0{.}02$ \\
		II-b  & $log(\mu_i)$&$\beta_0 = 3$, $\beta_1 = 2$, $\beta_2 = 1$ &  $x_{i1}\sim U(0,1); x_{i2}\sim U(0,1)$&$(20{.}085 ; 403{.}429)$ & $0{.}01$ \\
		III-b & $log(\mu_i)$& $\beta_0 = 3$, $\beta_1 = 2$, $\beta_2 = 1$ &  $x_{i1}\sim U(0,1); x_{i2}\sim U(0,1)$& $(20{.}085 ; 403{.}429)$ & $0{.}03$ \\
		IV-b  & $log(\mu_i)$& $\beta_0 = -3$, $\beta_1 = 1{.}5$, $\beta_2 = 1$ &  $x_{i1}\sim U(0,1); x_{i2}\sim U(0,1)$& $(0{.}049; 0{.}606)$ & $0{.}50$ \\
		V-b   & $log(\mu_i)$& $\beta_0 = 3$, $\beta_1 = 2$, $\beta_2 = 1$ &  $x_{i1}\sim NO(0.5,0.25^2); x_{i2}\sim IG(0.4,2)$& $(20{.}085 ; 403{.}429)$ & $0{.}02$ \\
		VI-b  &  $log(\mu_i)$&$\beta_0 = 3$, $\beta_1 = 2$, $\beta_2 = 1$ &   $x_{i1}\sim NO(0.5,0.25^2); x_{i2}\sim GA(0.4,1)$& $(7{.}400 ; 1892{.}000)$ & $0{.}02$ \\
		VII-b & $\frac{1}{\mu_i^2}$& $\beta_{0}=0{.}000006$, $\beta_1=0{.}002$, $\beta_2=0.001$ &  $x_{i1}\sim U(0,1); x_{i2}\sim U(0,1)$& $(18{.}24 ; 408{.}25)$ & $0{.}02$ \\
		\hline
	\end{tabular}%
	\label{tab:2}%
\end{table}%

Tables \ref{cen1ga} and \ref{tab:cen1} present the simulation results considering $n=15$ in the Scenario I-a for gamma regression model and in the Scenario I-b for inverse Gaussian regression model, respectively.     
 The means of \aqu\, and \pea\ are very close to zero, but the other residuals have means slightly lower than zero. 
 For all residuals, the variance is close to one, but none of the residuals have a variance close to one for all $15$ observations.
 The skewness coefficient is considerably greater than zero for \pea\ specially for inverse Gaussian regression model, very close to zero for  \wil\ and slightly lower than zero for the other residuals. 
 The excess kurtosis coefficients are similar and not too close to zero for the five residuals. 
 The residual \aqu\, has smaller AD statistic than the other residuals in 10 out of 15 observations for gamma regression model and 14 out of 15 observations for inverse Gaussian regression model.
 The AD statistic is smaller for the adjusted quantile residual, than that for the other residuals, because the mean of \aqu\, is closer to zero than that of \dev, \wil\, and \aasc\, and the skewness is closer to zero than that of \pea. 
\begin{landscape}
% Table generated by Excel2LaTeX from sheet 'cenario1'
\begin{table}[htbp]
	\centering
	\tabcolsep=0.07cm    
	\caption{Simulation results for Scenario I-a - gamma regression model.}
	\scalebox{0.8}{
		% Table generated by Excel2LaTeX from sheet 'Planilha4'
		\begin{tabular}{rrrrrrrrrrrrrrrrrrrrrrrrrrrrrrr}
			\cline{1-14}\cline{15-31}    i     & $\mu$  & \multicolumn{5} {c} {Mean} & & \multicolumn{5} {c} {Variance} & & \multicolumn{5} {c} {Skewness} & & \multicolumn{5} {c} {Kurtosis} & & \multicolumn{5} {c} {A-D} \\ 
			\cline{3-7}\cline{9-13}\cline{15-19}\cline{21-25}\cline{27-31}          &       & \dev  & \pea  & \aqu  & \aasc &  \wil  & & \dev  & \pea  & \aqu  & \aasc & \wil  &       & \dev  & \pea  & \aqu  & \aasc & \wil  &       & \dev  & \pea  & \aqu  & \aasc & \wil  &       & \dev  & \pea  & \aqu  & \aasc & \wil \\
		\hline
		\multicolumn{1}{c}{1} & 67.742 & -0.045 & -0.015 & -0.009 & -0.045 & -0.040 &       & 1.032 & 1.026 & 1.031 & 1.032 & 1.031 &       & -0.050 & 0.093 & -0.050 & -0.049 & -0.024 &       & -0.400 & -0.412 & -0.398 & -0.401 & -0.404 &       & 7.600 & 5.563 & 3.408 & 7.589 & 6.934 \\
		\multicolumn{1}{c}{2} & 100.306 & -0.011 & 0.018 & 0.024 & -0.011 & -0.007 &       & 0.984 & 0.984 & 0.984 & 0.983 & 0.984 &       & -0.029 & 0.109 & -0.029 & -0.029 & -0.009 &       & -0.463 & -0.452 & -0.462 & -0.464 & -0.463 &       & 2.751 & 3.512 & 4.191 & 2.751 & 2.596 \\
		\multicolumn{1}{c}{3} & 215.703 & -0.047 & -0.017 & -0.011 & -0.047 & -0.041 &       & 1.022 & 1.018 & 1.022 & 1.022 & 1.022 &       & -0.011 & 0.125 & -0.011 & -0.011 & 0.015 &       & -0.454 & -0.440 & -0.454 & -0.455 & -0.454 &       & 8.819 & 6.580 & 3.924 & 8.810 & 8.071 \\
		\multicolumn{1}{c}{4} & 115.585 & -0.032 & -0.001 & 0.003 & -0.032 & -0.028 &       & 0.986 & 0.985 & 0.985 & 0.985 & 0.986 &       & -0.006 & 0.136 & -0.006 & -0.006 & 0.009 &       & -0.446 & -0.441 & -0.444 & -0.447 & -0.447 &       & 5.349 & 5.357 & 2.781 & 5.349 & 5.130 \\
		\multicolumn{1}{c}{5} & 131.633 & -0.031 & -0.001 & 0.004 & -0.031 & -0.028 &       & 0.967 & 0.962 & 0.967 & 0.967 & 0.967 &       & -0.088 & 0.055 & -0.089 & -0.088 & -0.075 &       & -0.422 & -0.434 & -0.420 & -0.423 & -0.425 &       & 3.824 & 2.365 & 2.733 & 3.827 & 3.507 \\
		\multicolumn{1}{c}{6} & 200.681 & -0.012 & 0.017 & 0.024 & -0.012 & -0.006 &       & 0.978 & 0.975 & 0.978 & 0.978 & 0.978 &       & -0.088 & 0.056 & -0.088 & -0.088 & -0.062 &       & -0.336 & -0.366 & -0.335 & -0.337 & -0.344 &       & 1.384 & 1.884 & 3.543 & 1.386 & 1.177 \\
		\multicolumn{1}{c}{7} & 33.844 & -0.020 & 0.007 & 0.019 & -0.020 & -0.012 &       & 1.002 & 0.997 & 1.002 & 1.002 & 1.001 &       & -0.104 & 0.025 & -0.104 & -0.104 & -0.065 &       & -0.392 & -0.421 & -0.393 & -0.393 & -0.404 &       & 2.718 & 2.108 & 4.099 & 2.715 & 2.163 \\
		\multicolumn{1}{c}{8} & 80.061 & -0.025 & 0.005 & 0.011 & -0.025 & -0.020 &       & 0.993 & 0.991 & 0.993 & 0.993 & 0.993 &       & -0.032 & 0.113 & -0.032 & -0.032 & -0.009 &       & -0.372 & -0.368 & -0.371 & -0.373 & -0.374 &       & 2.631 & 2.484 & 1.759 & 2.629 & 2.325 \\
		\multicolumn{1}{c}{9} & 48.232 & -0.022 & 0.004 & 0.018 & -0.022 & -0.013 &       & 0.991 & 0.988 & 0.990 & 0.990 & 0.990 &       & -0.059 & 0.066 & -0.058 & -0.059 & -0.016 &       & -0.433 & -0.445 & -0.433 & -0.434 & -0.440 &       & 3.077 & 2.835 & 3.506 & 3.076 & 2.618 \\
		\multicolumn{1}{c}{10} & 52.030 & -0.029 & -0.002 & 0.009 & -0.029 & -0.022 &       & 0.986 & 0.982 & 0.985 & 0.986 & 0.985 &       & -0.041 & 0.094 & -0.041 & -0.041 & -0.006 &       & -0.374 & -0.363 & -0.374 & -0.375 & -0.375 &       & 3.213 & 1.865 & 2.006 & 3.212 & 2.501 \\
		\multicolumn{1}{c}{11} & 48.729 & -0.039 & -0.009 & -0.004 & -0.039 & -0.034 &       & 1.020 & 1.014 & 1.020 & 1.020 & 1.020 &       & -0.061 & 0.080 & -0.061 & -0.061 & -0.038 &       & -0.424 & -0.439 & -0.423 & -0.425 & -0.429 &       & 6.634 & 4.850 & 3.670 & 6.627 & 6.053 \\
		\multicolumn{1}{c}{12} & 114.403 & -0.028 & -0.002 & 0.013 & -0.028 & -0.018 &       & 1.033 & 1.031 & 1.032 & 1.032 & 1.032 &       & -0.012 & 0.110 & -0.012 & -0.012 & 0.034 &       & -0.433 & -0.428 & -0.433 & -0.434 & -0.434 &       & 6.159 & 5.604 & 4.688 & 6.153 & 5.574 \\
		\multicolumn{1}{c}{13} & 194.271 & -0.046 & -0.015 & -0.010 & -0.046 & -0.040 &       & 1.026 & 1.021 & 1.025 & 1.025 & 1.025 &       & -0.018 & 0.126 & -0.018 & -0.018 & 0.007 &       & -0.403 & -0.402 & -0.403 & -0.404 & -0.406 &       & 8.512 & 6.922 & 3.815 & 8.504 & 7.941 \\
		\multicolumn{1}{c}{14} & 103.731 & -0.033 & -0.001 & 0.001 & -0.033 & -0.030 &       & 0.999 & 0.997 & 0.998 & 0.998 & 0.999 &       & -0.022 & 0.127 & -0.022 & -0.022 & -0.011 &       & -0.430 & -0.402 & -0.427 & -0.431 & -0.429 &       & 4.591 & 3.504 & 2.320 & 4.588 & 4.358 \\
		\multicolumn{1}{c}{15} & 208.921 & -0.017 & 0.013 & 0.021 & -0.017 & -0.010 &       & 1.043 & 1.042 & 1.043 & 1.043 & 1.043 &       & -0.045 & 0.086 & -0.046 & -0.045 & -0.014 &       & -0.471 & -0.462 & -0.470 & -0.472 & -0.472 &       & 6.092 & 6.058 & 7.130 & 6.081 & 5.729 \\
		\hline
		& Mean  & -0.029 & 0.000 & 0.007 & -0.029 & -0.023 &       & 1.004 & 1.001 & 1.004 & 1.004 & 1.004 &       & -0.045 & 0.093 & -0.045 & -0.044 & -0.018 &       & -0.417 & -0.418 & -0.416 & -0.418 & -0.420 &       & 4.890 & 4.099 & 3.572 & 4.887 & 4.445 \\
		& SD    & 0.012 & 0.011 & 0.012 & 0.012 & 0.012 &       & 0.023 & 0.023 & 0.023 & 0.023 & 0.023 &       & 0.031 & 0.032 & 0.031 & 0.031 & 0.031 &       & 0.037 & 0.032 & 0.037 & 0.037 & 0.036 &       & 2.333 & 1.818 & 1.294 & 2.330 & 2.226 \\
		\hline
		\end{tabular}%
	}
	\label{cen1ga}%
\end{table}
\end{landscape}

	\begin{landscape}
	% Table generated by Excel2LaTeX from sheet 'cenario1'
	\begin{table}[htbp]
		\centering
		\tabcolsep=0.07cm    
		\caption{Simulation results for Scenario I-b - inverse Gaussian regression model.}
		\scalebox{0.8}{
			\begin{tabular}{rrrrrrrrrrrrrrrrrrrrrrrrrrrrrrr}
				\cline{1-14}\cline{15-31}    i     & $\mu$  & \multicolumn{5} {c} {Mean} & & \multicolumn{5} {c} {Variance} & & \multicolumn{5} {c} {Skewness} & & \multicolumn{5} {c} {Kurtosis} & & \multicolumn{5} {c} {A-D} \\ 
				\cline{3-7}\cline{9-13}\cline{15-19}\cline{21-25}\cline{27-31}          &       & \dev  & \pea  & \aqu  & \aasc &  \wil  & & \dev  & \pea  & \aqu  & \aasc & \wil  &       & \dev  & \pea  & \aqu  & \aasc & \wil  &       & \dev  & \pea  & \aqu  & \aasc & \wil  &       & \dev  & \pea  & \aqu  & \aasc & \wil \\
				% Table generated by Excel2LaTeX from sheet 'Planilha1'
					\cline{1-7}\cline{9-13}\cline{15-19}\cline{21-25}\cline{27-31}1     & 67.742 & -0.080 & -0.006 & 0.008 & -0.079 & -0.067 &       & 1.028 & 1.019 & 1.034 & 1.023 & 1.027 &       & -0.058 & 0.278 & -0.060 & -0.056 & 0.001 &       & -0.430 & -0.382 & -0.424 & -0.442 & -0.437 &       & 18.165 & 12.859 & 5.000 & 18.038 & 15.419 \\
					2     & 100.306 & -0.096 & -0.008 & 0.007 & -0.096 & -0.084 &       & 0.994 & 0.975 & 1.003 & 0.987 & 0.992 &       & -0.121 & 0.310 & -0.123 & -0.117 & -0.062 &       & -0.312 & -0.289 & -0.307 & -0.335 & -0.330 &       & 20.418 & 12.314 & 2.700 & 20.318 & 17.068 \\
					3     & 215.703 & -0.107 & 0.029 & 0.042 & -0.105 & -0.089 &       & 1.051 & 1.032 & 1.072 & 1.034 & 1.050 &       & -0.166 & 0.432 & -0.171 & -0.154 & -0.089 &       & -0.336 & -0.338 & -0.321 & -0.384 & -0.373 &       & 29.403 & 26.595 & 15.513 & 28.836 & 24.399 \\
					4     & 115.585 & -0.089 & 0.013 & 0.020 & -0.088 & -0.077 &       & 1.041 & 1.036 & 1.053 & 1.032 & 1.042 &       & -0.093 & 0.374 & -0.093 & -0.088 & -0.040 &       & -0.380 & -0.305 & -0.369 & -0.403 & -0.391 &       & 20.837 & 18.148 & 6.654 & 20.560 & 18.235 \\
					5     & 131.633 & -0.080 & 0.024 & 0.036 & -0.079 & -0.069 &       & 1.002 & 0.993 & 1.014 & 0.992 & 1.002 &       & -0.108 & 0.356 & -0.111 & -0.103 & -0.062 &       & -0.434 & -0.380 & -0.426 & -0.458 & -0.446 &       & 17.200 & 16.717 & 9.541 & 17.093 & 14.858 \\
					6     & 200.681 & -0.122 & 0.005 & 0.021 & -0.121 & -0.106 &       & 1.016 & 0.991 & 1.035 & 1.000 & 1.014 &       & -0.168 & 0.435 & -0.172 & -0.158 & -0.091 &       & -0.256 & -0.204 & -0.240 & -0.300 & -0.287 &       & 32.684 & 23.010 & 6.316 & 32.316 & 27.199 \\
					7     & 33.844 & -0.004 & 0.037 & 0.074 & -0.004 & 0.015 &       & 1.008 & 1.002 & 1.012 & 1.007 & 1.005 &       & -0.168 & 0.025 & -0.166 & -0.167 & -0.076 &       & -0.395 & -0.480 & -0.395 & -0.400 & -0.444 &       & 4.948 & 6.775 & 22.802 & 4.936 & 4.828 \\
					8     & 80.061 & -0.079 & 0.001 & 0.014 & -0.079 & -0.067 &       & 1.020 & 1.012 & 1.028 & 1.014 & 1.019 &       & -0.077 & 0.301 & -0.077 & -0.074 & -0.019 &       & -0.384 & -0.337 & -0.379 & -0.400 & -0.395 &       & 17.409 & 13.792 & 5.174 & 17.304 & 14.916 \\
					9     & 48.232 & -0.080 & -0.030 & 0.010 & -0.080 & -0.058 &       & 0.993 & 0.973 & 0.996 & 0.990 & 0.985 &       & -0.129 & 0.110 & -0.130 & -0.127 & -0.024 &       & -0.378 & -0.416 & -0.377 & -0.384 & -0.408 &       & 14.443 & 6.589 & 3.868 & 14.414 & 9.515 \\
					10    & 52.030 & -0.086 & -0.028 & -0.001 & -0.085 & -0.067 &       & 0.996 & 0.983 & 1.001 & 0.993 & 0.992 &       & -0.079 & 0.197 & -0.080 & -0.078 & 0.008 &       & -0.378 & -0.366 & -0.372 & -0.385 & -0.389 &       & 17.701 & 10.138 & 2.527 & 17.669 & 13.627 \\
					11    & 48.729 & -0.060 & 0.002 & 0.013 & -0.060 & -0.050 &       & 1.016 & 1.007 & 1.020 & 1.012 & 1.015 &       & -0.094 & 0.201 & -0.095 & -0.093 & -0.046 &       & -0.364 & -0.363 & -0.361 & -0.374 & -0.375 &       & 10.242 & 7.378 & 4.495 & 10.192 & 8.554 \\
					12    & 114.403 & -0.106 & -0.017 & 0.016 & -0.105 & -0.080 &       & 1.019 & 0.988 & 1.028 & 1.011 & 1.011 &       & -0.149 & 0.265 & -0.149 & -0.144 & -0.030 &       & -0.349 & -0.380 & -0.341 & -0.372 & -0.390 &       & 25.359 & 12.683 & 6.055 & 25.137 & 17.849 \\
					13    & 194.271 & -0.144 & -0.019 & -0.003 & -0.142 & -0.128 &       & 1.007 & 0.969 & 1.024 & 0.992 & 1.003 &       & -0.148 & 0.443 & -0.150 & -0.137 & -0.074 &       & -0.323 & -0.208 & -0.309 & -0.362 & -0.342 &       & 45.361 & 25.792 & 3.474 & 44.880 & 38.776 \\
					14    & 103.731 & -0.083 & 0.014 & 0.019 & -0.083 & -0.075 &       & 1.028 & 1.026 & 1.038 & 1.020 & 1.029 &       & -0.067 & 0.376 & -0.066 & -0.064 & -0.031 &       & -0.421 & -0.308 & -0.412 & -0.439 & -0.424 &       & 18.672 & 16.773 & 5.574 & 18.437 & 16.877 \\
					15    & 208.921 & -0.147 & -0.018 & 0.001 & -0.145 & -0.128 &       & 1.016 & 0.976 & 1.036 & 1.001 & 1.013 &       & -0.153 & 0.426 & -0.156 & -0.142 & -0.068 &       & -0.361 & -0.308 & -0.351 & -0.398 & -0.389 &       & 48.182 & 28.817 & 5.324 & 47.592 & 40.397 \\
				\hline
			& Mean  & -0.091 & 0.000 & 0.018 & -0.090 & -0.075 &       & 1.016 & 0.999 & 1.026 & 1.007 & 1.013 &       & -0.119 & 0.302 & -0.120 & -0.113 & -0.047 &       & -0.367 & -0.338 & -0.359 & -0.389 & -0.388 &       & 22.735 & 15.892 & 7.001 & 22.515 & 18.834 \\
			& SD    & 0.034 & 0.021 & 0.020 & 0.034 & 0.034 &       & 0.017 & 0.022 & 0.020 & 0.015 & 0.018 &       & 0.039 & 0.125 & 0.040 & 0.037 & 0.031 &       & 0.047 & 0.072 & 0.050 & 0.040 & 0.044 &       & 11.898 & 7.313 & 5.407 & 11.722 & 10.108 \\
			\hline
				\end{tabular}%
		}
		\label{tab:cen1}%
	\end{table}%
\end{landscape}
Table \ref{tab:gama} presents the average values for the mean, variance, skewness and excess kurtosis for all residuals and both sample sizes for gamma regression model. Changes in the range of $\mu$, in the distribution of the covariates and in the link function does not seem to affect considerably the distribution of the residuals, since the results of Scenarios IV-a to VII-a are very similar to that of Scenario I-a. However, the distributions of the five residuals are worse approximated by the standard normal distribution when variance increases. The skewness coefficients are far from zero in Scenario III-a than in Scenario I-a and, except for \pea\,, the mean and variance are also far from zero and one, respectively. As a consequence, the AD statistics increase substantially from Scenario I-a to Scenario III-a, but the rise is smaller for \aqu\,.
 
In general, when sample size increases from $n=15$ to $n=50$, all residuals's distributions are better approximated by the standard normal distribution. 
In all scenarios, the absolute value of the average skewness coefficient reduces considerably, except for the residual \pea\,, in which the average skewness coefficient rises slightly.
Additionally, the absolute average excess kurtosis coefficient also decreases in all scenarios for all residuals, except for \pea\, in Scenario III-a. 
Finally, in six out of seven scenarios, the AD statistic reduces for all residuals, but the reduction is greater for \aqu\,.

Table \ref{adga} summarizes the results of the AD statistic in all scenarios for $n=15$ and $n=50$ considering the gamma regression model.
When $n=50$ the mean, the standard deviation and the maximum of the AD statistic are lower for \aqu\, than for other residuals. 
When sample size is 15, the sample mean of the AD statistic is also smaller for  \aqu\, than for other residuals, except in Scenario IV-a in which \pea\, and \aqu\, have similar sample mean of the AD statistic. On the other hand, the standard deviation and the maximum are the lowest for \aqu\, in some scenarios and for \pea\, in others.

Tables \ref{tab:ig} and \ref{adig} present the average of the sample moments for the residuals and a descriptive summary of the results of the AD statistic for the inverse Gaussian regression model. 
For all residuals, in the most of the scenarios, the average values of the AD statistic are higher for the inverse Gaussian regression model than for the gamma regression model. A plausible explanation for this result is that, for the same mean and variance, the inverse Gaussian distribution presents a greater skewness than the gamma distribution. In general, the standard deviation of AD statistic are also higher for the inverse Gaussian regression model. Residual \aqu\, presents lower mean of the AD statistic than the other residuals for all scenarios and both sample sizes. Additionally, the differences between the mean AD statistic for \aqu\, and for the other residuals are greater for the inverse Gaussian regression model. Finally, except for Scenario V-a when $n=15$, the standard deviation of the AD statistic is lower for \aqu\, than for the other residuals.

		% Table generated by Excel2LaTeX from sheet 'Planilha2'

		\begin{table}[htbp]
			
			\tabcolsep=0.1cm    
			\centering
			\caption{Distributional measures for the residual in Gamma regression model.}
			\scalebox{.8}{
				\begin{tabular}{crrrrrrrrrrrrrrr}
					\hline
					&       & \multicolumn{2}{c}{\dev} & \multicolumn{1}{c}{} & \multicolumn{2}{c}{\pea} & \multicolumn{1}{c}{} & \multicolumn{2}{c}{\aqu} & \multicolumn{1}{c}{} & \multicolumn{2}{c}{\aasc} & \multicolumn{1}{c}{} & \multicolumn{2}{c}{\wil} \\
					\cline{3-4}\cline{6-7}\cline{9-10}\cline{12-13}\cline{15-16}    Scenarios &       & n=15  & n=50  &       & n=15  & n=50  &       & n=15  & n=50  &       & n=15  & n=50  &       & n=15  & n=50 \\
					\hline
					& Mean  & -0.03 & -0.03 &       & 0.00  & 0.00  &       & 0.01  & 0.00  &       & -0.03 & -0.03 &       & -0.02 & -0.03 \\
					& Variance    & 1.00  & 1.00  &       & 1.00  & 1.00  &       & 1.00  & 1.00  &       & 1.00  & 1.00  &       & 1.00  & 1.00 \\
					I-a     & Skewness & -0.04 & -0.02 &       & 0.09  & 0.16  &       & -0.04 & -0.02 &       & -0.04 & -0.02 &       & -0.02 & -0.01 \\
					& Kurtosis & -0.42 & -0.11 &       & -0.42 & -0.08 &       & -0.42 & -0.11 &       & -0.42 & -0.11 &       & -0.42 & -0.11 \\
					& AD Statistics & 4.89  & 3.43  &       & 4.10  & 2.84  &       & 3.57  & 1.13  &       & 4.89  & 3.43  &       & 4.45  & 3.26 \\
					
					\hline
					& Mean  & -0.01 & -0.02 &       & 0.00  & 0.00  &       & 0.00  & 0.00  &       & -0.01 & -0.02 &       & -0.01 & -0.02 \\
					& Variance    & 1.00  & 1.00  &       & 1.00  & 1.00  &       & 1.00  & 1.00  &       & 1.00  & 1.00  &       & 1.00  & 1.00 \\
					II-a    & Skewness & -0.01 & -0.01 &       & 0.06  & 0.09  &       & -0.01 & -0.01 &       & -0.01 & -0.01 &       & 0.00  & 0.00 \\
					& Kurtosis & -0.42 & -0.13 &       & -0.42 & -0.12 &       & -0.42 & -0.13 &       & -0.42 & -0.13 &       & -0.43 & -0.13 \\
					& AD Statistics & 4.05  & 1.64  &       & 3.94  & 1.52  &       & 3.62  & 1.03  &       & 4.05  & 1.64  &       & 3.94  & 1.60 \\
					\hline
					& Mean  & -0.15 & -0.16 &       & 0.00  & 0.00  &       & 0.03  & 0.01  &       & -0.15 & -0.16 &       & -0.13 & -0.16 \\
					& Variance    & 1.08  & 1.04  &       & 1.00  & 1.00  &       & 1.07  & 1.03  &       & 1.07  & 1.03  &       & 1.06  & 1.04 \\

					III-a   & Skewness & -0.22 & -0.09 &       & 0.46  & 0.79  &       & -0.26 & -0.10 &       & -0.22 & -0.07 &       & -0.09 & -0.04 \\
					& Kurtosis & -0.24 & -0.10 &       & -0.30 & 0.62  &       & -0.02 & -0.09 &       & -0.22 & -0.16 &       & -0.32 & -0.10 \\
					& AD Statistics & 57.42 & 62.34 &       & 35.35 & 45.69 &       & 14.41 & 2.36  &       & 56.75 & 61.41 &       & 44.25 & 57.93 \\
					\hline
					
					& Mean  & -0.03 & -0.03 &       & 0.00  & 0.00  &       & 0.01  & 0.00  &       & -0.03 & -0.03 &       & -0.02 & -0.03 \\
					& Variance    & 1.00  & 1.00  &       & 1.00  & 1.00  &       & 1.00  & 1.00  &       & 1.00  & 1.00  &       & 1.00  & 1.00 \\
					IV-a    & Skewness & -0.06 & -0.01 &       & 0.08  & 0.17  &       & -0.06 & -0.01 &       & -0.06 & -0.01 &       & -0.03 & 0.00 \\
					& Kurtosis & -0.41 & -0.12 &       & -0.42 & -0.09 &       & -0.41 & -0.12 &       & -0.41 & -0.12 &       & -0.42 & -0.12 \\
					& AD Statistics & 4.97  & 3.57  &       & 4.00  & 3.08  &       & 4.01  & 1.17  &       & 4.97  & 3.57  &       & 4.53  & 3.41 \\
					\hline
					& Mean  & -0.03 & -0.03 &       & 0.00  & 0.00  &       & 0.01  & 0.00  &       & -0.03 & -0.03 &       & -0.02 & -0.03 \\
					& Variance    & 1.00  & 1.00  &       & 1.00  & 1.00  &       & 1.00  & 1.00  &       & 1.00  & 1.00  &       & 1.00  & 1.00 \\
					V-a   & Skewness & -0.03 & -0.02 &       & 0.11  & 0.16  &       & -0.03 & -0.02 &       & -0.03 & -0.02 &       & 0.00  & -0.01 \\
					& Kurtosis & -0.43 & -0.12 &       & -0.42 & -0.09 &       & -0.43 & -0.12 &       & -0.43 & -0.13 &       & -0.43 & -0.12 \\
					& AD Statistics & 5.44  & 3.60  &       & 4.83  & 2.94  &       & 3.97  & 1.48  &       & 5.43  & 3.60  &       & 5.04  & 3.44 \\
					\hline
					& Mean  & -0.03 & -0.03 &       & 0.00  & 0.00  &       & 0.01  & 0.00  &       & -0.03 & -0.03 &       & -0.02 & -0.03 \\
					& Variance    & 1.00  & 1.00  &       & 1.00  & 1.00  &       & 1.00  & 1.00  &       & 1.00  & 1.00  &       & 1.00  & 1.00 \\
					VI-a    & Skewness & -0.04 & -0.02 &       & 0.10  & 0.16  &       & -0.04 & -0.02 &       & -0.04 & -0.02 &       & -0.01 & -0.01 \\
					& Kurtosis & -0.41 & -0.12 &       & -0.41 & -0.09 &       & -0.41 & -0.12 &       & -0.41 & -0.12 &       & -0.41 & -0.12 \\
					& AD Statistics & 4.83  & 3.50  &       & 4.22  & 2.92  &       & 3.18  & 1.15  &       & 4.82  & 3.50  &       & 4.34  & 3.32 \\
					\hline
					& Mean  & -0.03 & -0.03 &       & 0.00  & 0.00  &       & 0.01  & 0.00  &       & -0.03 & -0.03 &       & -0.02 & -0.03 \\
					& Variance    & 1.00  & 1.00  &       & 1.00  & 1.00  &       & 1.00  & 1.00  &       & 1.00  & 1.00  &       & 1.00  & 1.00 \\
					VII-a     & Skewness & -0.03 & -0.02 &       & 0.10  & 0.16  &       & -0.03 & -0.02 &       & -0.03 & -0.02 &       & -0.01 & -0.01 \\
					& Kurtosis & -0.40 & -0.11 &       & -0.40 & -0.08 &       & -0.40 & -0.11 &       & -0.41 & -0.11 &       & -0.41 & -0.11 \\
					& AD Statistics & 5.32  & 3.35  &       & 4.64  & 2.79  &       & 3.82  & 1.01  &       & 5.31  & 3.35  &       & 4.87  & 3.18 \\
					\hline
				\end{tabular}%
			}
			\label{tab:gama}%
		\end{table}%
\begin{landscape}
	
	\begin{table}
		\centering
		\caption{Comparison of the Anderson-Darling statistic for n = 15 and n = 50 - gamma regression model.}
		\tabcolsep=0.2cm    \scalebox{.6}{

\begin{tabular}{ccccrrrrrrr|rrrrrrrrrrr}
	\hline
	&       &       &       &       & \multicolumn{1}{c}{\textbf{Standard}} &       & \multicolumn{3}{c}{\textbf{Quantiles}} &       &       &       &       &       &       & \multicolumn{1}{c}{\textbf{Standard}} &       & \multicolumn{3}{c}{\textbf{Quantiles}} &  \\
	\cline{8-10}\cline{19-21}\textbf{Scenario} & \textbf{sigma} & \textbf{n} & \textbf{Resíduos} & \multicolumn{1}{c}{\textbf{Mean}} & \multicolumn{1}{c}{\textbf{Deviantion}} & \multicolumn{1}{c}{\textbf{Minimum}} & \multicolumn{1}{c}{\textbf{Q1}} & \multicolumn{1}{c}{\textbf{Q2}} & \multicolumn{1}{c}{\textbf{Q3}} & \multicolumn{1}{c|}{\textbf{Maximum}} & \multicolumn{1}{c}{\textbf{Scenario}} & \multicolumn{1}{c}{\textbf{sigma}} & \multicolumn{1}{c}{\textbf{n}} & \multicolumn{1}{c}{\textbf{Resíduos}} & \multicolumn{1}{c}{\textbf{Mean}} & \multicolumn{1}{c}{\textbf{Deviantion}} & \multicolumn{1}{c}{\textbf{Minimum}} & \multicolumn{1}{c}{\textbf{Q1}} & \multicolumn{1}{c}{\textbf{Q2}} & \multicolumn{1}{c}{\textbf{Q3}} & \multicolumn{1}{c}{\textbf{Maximum}} \\
	\hline
	&       &       & \dev  & 4.89  & 2.33  & 1.38  & 2.91  & 4.59  & 6.52  & 8.82  &       &       &       & \multicolumn{1}{c}{\dev} & 5.44  & 1.78  & 2.21  & 4.35  & 5.62  & 6.40  & 8.37 \\
	&       &       & \pea  & 4.10  & 1.82  & 1.86  & 2.42  & 3.51  & 6.52  & 6.92  &       &       &       & \multicolumn{1}{c}{\pea} & 4.83  & 0.98  & 3.24  & 4.30  & 4.60  & 5.43  & 6.75 \\
	\textbf{I-a} & \textbf{0.1} & \textbf{15} & \aqu  & 3.57  & 1.29  & 1.76  & 2.76  & 3.54  & 6.52  & 7.13  & \multicolumn{1}{c}{\textbf{V-a}} & \multicolumn{1}{c}{\textbf{0.1}} & \multicolumn{1}{c}{\textbf{15}} & \multicolumn{1}{c}{\aqu} & 3.97  & 2.23  & 1.68  & 2.55  & 3.62  & 4.55  & 10.27 \\
	&       &       & \aasc & 4.89  & 2.33  & 1.39  & 2.91  & 4.59  & 6.52  & 8.81  &       &       &       & \multicolumn{1}{c}{\aasc} & 5.43  & 1.77  & 2.21  & 4.35  & 5.61  & 6.40  & 8.36 \\
	&       &       & \wil  & 4.45  & 2.23  & 1.18  & 2.55  & 4.36  & 6.52  & 8.07  &       &       &       & \multicolumn{1}{c}{\wil} & 5.04  & 1.54  & 2.25  & 4.06  & 5.06  & 5.83  & 7.83 \\
	\hline
	&       &       & \dev  & 3.43  & 2.34  & 0.49  & 1.55  & 2.82  & 5.32  & 10.92 &       &       &       & \multicolumn{1}{c}{\dev} & 3.60  & 2.68  & 0.40  & 1.86  & 2.64  & 4.79  & 11.72 \\
	&       &       & \pea  & 2.84  & 1.55  & 0.55  & 1.62  & 2.47  & 3.82  & 7.45  &       &       &       & \multicolumn{1}{c}{\pea} & 2.94  & 1.58  & 0.46  & 1.71  & 2.54  & 3.53  & 7.43 \\
	\textbf{I-a} & \textbf{0.1} & \textbf{50} & \aqu  & 1.13  & 0.73  & 0.26  & 0.66  & 0.90  & 1.28  & 3.25  & \multicolumn{1}{c}{\textbf{V-a}} & \multicolumn{1}{c}{\textbf{0.1}} & \multicolumn{1}{c}{\textbf{50}} & \multicolumn{1}{c}{\aqu} & 1.48  & 1.45  & 0.33  & 0.69  & 0.94  & 1.61  & 9.27 \\
	&       &       & \aasc & 3.43  & 2.34  & 0.49  & 1.55  & 2.82  & 5.31  & 10.92 &       &       &       & \multicolumn{1}{c}{\aasc} & 3.60  & 2.68  & 0.39  & 1.85  & 2.63  & 4.79  & 11.71 \\
	&       &       & \wil  & 3.26  & 2.27  & 0.45  & 1.44  & 2.62  & 4.98  & 10.69 &       &       &       & \multicolumn{1}{c}{\wil} & 3.44  & 2.60  & 0.36  & 1.75  & 2.50  & 4.39  & 11.57 \\
	\hline
	&       &       & \dev  & 4.97  & 1.89  & 1.70  & 3.80  & 4.35  & 6.58  & 8.30  &       &       &       & \multicolumn{1}{c}{\dev} & 4.83  & 1.77  & 2.23  & 3.64  & 4.52  & 6.27  & 8.22 \\
	&       &       & \pea  & 4.00  & 1.51  & 1.62  & 3.04  & 3.94  & 5.07  & 7.52  &       &       &       & \multicolumn{1}{c}{\pea} & 4.22  & 1.13  & 1.94  & 3.78  & 4.22  & 4.67  & 6.44 \\
	\textbf{II-a} & \textbf{0.1} & \textbf{15} & \aqu  & 4.01  & 1.82  & 2.01  & 2.72  & 3.55  & 4.64  & 9.14  & \multicolumn{1}{c}{\textbf{VI-a}} & \multicolumn{1}{c}{\textbf{0.1}} & \multicolumn{1}{c}{\textbf{15}} & \multicolumn{1}{c}{\aqu} & 3.18  & 1.08  & 1.44  & 2.51  & 3.32  & 3.67  & 5.39 \\
	&       &       & \aasc & 4.97  & 1.89  & 1.70  & 3.79  & 4.34  & 6.58  & 8.29  &       &       &       & \multicolumn{1}{c}{\aasc} & 4.82  & 1.77  & 2.23  & 3.64  & 4.52  & 6.26  & 8.22 \\
	&       &       & \wil  & 4.53  & 1.84  & 1.31  & 3.28  & 4.05  & 5.88  & 8.09  &       &       &       & \multicolumn{1}{c}{\wil} & 4.34  & 1.47  & 1.86  & 3.41  & 4.08  & 5.25  & 7.33 \\
	\hline
	&       &       & \dev  & 3.57  & 2.53  & 0.32  & 1.92  & 2.91  & 4.40  & 11.85 &       &       &       & \multicolumn{1}{c}{\dev} & 3.50  & 2.58  & 0.39  & 1.71  & 2.70  & 4.98  & 10.83 \\
	&       &       & \pea  & 3.08  & 1.69  & 0.84  & 2.02  & 2.76  & 3.40  & 8.57  &       &       &       & \multicolumn{1}{c}{\pea} & 2.92  & 1.58  & 1.12  & 1.80  & 2.38  & 3.59  & 7.40 \\
	\textbf{II-a} & \textbf{0.1} & \textbf{50} & \aqu  & 1.17  & 0.90  & 0.22  & 0.57  & 0.77  & 1.52  & 3.89  & \multicolumn{1}{c}{\textbf{VI-a}} & \multicolumn{1}{c}{\textbf{0.1}} & \multicolumn{1}{c}{\textbf{50}} & \multicolumn{1}{c}{\aqu} & 1.15  & 0.74  & 0.19  & 0.62  & 0.88  & 1.46  & 3.27 \\
	&       &       & \aasc & 3.57  & 2.53  & 0.32  & 1.91  & 2.91  & 4.40  & 11.84 &       &       &       & \multicolumn{1}{c}{\aasc} & 3.50  & 2.58  & 0.39  & 1.71  & 2.70  & 4.98  & 10.82 \\
	&       &       & \wil  & 3.41  & 2.47  & 0.32  & 1.84  & 2.73  & 4.21  & 11.66 &       &       &       & \multicolumn{1}{c}{\wil} & 3.32  & 2.46  & 0.33  & 1.61  & 2.61  & 4.62  & 10.68 \\
	\hline
	&       &       & \dev  & 4.05  & 2.36  & 1.19  & 2.47  & 3.66  & 4.71  & 10.98 &       &       &       & \multicolumn{1}{c}{\dev} & 5.32  & 2.63  & 2.61  & 3.43  & 4.47  & 6.74  & 10.97 \\
	&       &       & \pea  & 3.94  & 1.98  & 1.18  & 2.59  & 3.76  & 4.31  & 9.79  &       &       &       & \multicolumn{1}{c}{\pea} & 4.64  & 2.01  & 2.12  & 3.19  & 3.62  & 6.03  & 8.67 \\
	\textbf{III-a} & \textbf{0.05} & \textbf{15} & \aqu  & 3.62  & 1.97  & 0.61  & 2.59  & 3.37  & 3.96  & 9.06  & \multicolumn{1}{c}{\textbf{VII-a}} & \multicolumn{1}{c}{\textbf{0.1}} & \multicolumn{1}{c}{\textbf{15}} & \multicolumn{1}{c}{\aqu} & 3.82  & 2.11  & 1.49  & 2.35  & 3.32  & 4.45  & 9.31 \\
	&       &       & \aasc & 4.05  & 2.36  & 1.19  & 2.47  & 3.66  & 4.71  & 10.98 &       &       &       & \multicolumn{1}{c}{\aasc} & 5.31  & 2.62  & 2.61  & 3.43  & 4.47  & 6.74  & 10.96 \\
	&       &       & \wil  & 3.94  & 2.25  & 1.14  & 2.45  & 3.70  & 4.51  & 10.57 &       &       &       & \multicolumn{1}{c}{\wil} & 4.87  & 2.48  & 2.36  & 2.96  & 4.31  & 6.41  & 10.47 \\
	\hline
	&       &       & \dev  & 1.64  & 1.08  & 0.37  & 0.90  & 1.34  & 2.06  & 4.89  &       &       &       & \multicolumn{1}{c}{\dev} & 3.35  & 2.43  & 0.60  & 1.75  & 2.67  & 4.06  & 12.85 \\
	&       &       & \pea  & 1.52  & 0.73  & 0.39  & 1.00  & 1.36  & 1.77  & 3.65  &       &       &       & \multicolumn{1}{c}{\pea} & 2.79  & 1.48  & 0.90  & 1.77  & 2.55  & 3.18  & 9.11 \\
	\textbf{III-a} & \textbf{0.05} & \textbf{15} & \aqu  & 1.03  & 0.59  & 0.24  & 0.58  & 0.84  & 1.39  & 2.87  & \multicolumn{1}{c}{\textbf{VII-a}} & \multicolumn{1}{c}{\textbf{0.1}} & \multicolumn{1}{c}{\textbf{50}} & \multicolumn{1}{c}{\aqu} & 1.01  & 0.69  & 0.33  & 0.48  & 0.92  & 1.20  & 4.36 \\
	&       &       & \aasc & 1.64  & 1.08  & 0.36  & 0.90  & 1.34  & 2.06  & 4.89  &       &       &       & \multicolumn{1}{c}{\aasc} & 3.35  & 2.43  & 0.60  & 1.75  & 2.67  & 4.06  & 12.85 \\
	&       &       & \wil  & 1.60  & 1.06  & 0.39  & 0.89  & 1.30  & 2.03  & 4.82  &       &       &       & \multicolumn{1}{c}{\wil} & 3.18  & 2.31  & 0.58  & 1.56  & 2.58  & 3.96  & 12.54 \\
	\hline
	&       &       & \dev  & 57.42 & 29.40 & 25.41 & 34.77 & 46.37 & 76.42 & 107.52 &       &       &       &       &       &       &       &       &       &       &  \\
	&       &       & \pea  & 35.35 & 11.45 & 18.29 & 27.12 & 32.74 & 43.00 & 54.26 &       &       &       &       &       &       &       &       &       &       &  \\
	\textbf{IV-a} & \textbf{0.5} & \textbf{15} & \aqu  & 14.41 & 7.52  & 6.37  & 8.88  & 12.44 & 18.53 & 33.87 &       &       &       &       &       &       &       &       &       &       &  \\
	&       &       & \aasc & 56.75 & 29.28 & 25.15 & 33.86 & 45.77 & 75.51 & 106.78 &       &       &       &       &       &       &       &       &       &       &  \\
	&       &       & \wil  & 44.25 & 22.33 & 19.99 & 27.01 & 36.43 & 59.18 & 83.70 &       &       &       &       &       &       &       &       &       &       &  \\
	\cline{1-11}      &       &       & \dev  & 62.34 & 12.51 & 37.18 & 52.29 & 63.73 & 71.75 & 88.67 &       &       &       &       &       &       &       &       &       &       &  \\
	&       &       & \pea  & 45.69 & 6.69  & 31.47 & 41.53 & 44.93 & 50.16 & 63.62 &       &       &       &       &       &       &       &       &       &       &  \\
	\textbf{IV-a} & \textbf{0.5} & \textbf{50} & \aqu  & 2.36  & 1.74  & 0.52  & 1.24  & 1.79  & 2.81  & 7.11  &       &       &       &       &       &       &       &       &       &       &  \\
	&       &       & \aasc & 61.41 & 12.43 & 36.64 & 51.26 & 62.91 & 70.83 & 87.75 &       &       &       &       &       &       &       &       &       &       &  \\
	&       &       & \wil  & 57.93 & 11.79 & 33.03 & 49.04 & 59.14 & 66.71 & 84.92 &       &       &       &       &       &       &       &       &       &       &  \\
	\cline{1-11}   \end{tabular}%
			\label{adga}
		}
	\end{table}
\end{landscape}

	% Table generated by Excel2LaTeX from sheet 'Planilha3'
	\begin{table}[htbp]
		\centering
		\tabcolsep=0.1cm    
		\caption{Distributional measures for the residuals in inverse Gaussian regression model.}
		\scalebox{0.8}{
			\begin{tabular}{crrrrrrrrrrrrrrr}
				\hline
				&       & \multicolumn{2}{c}{\dev} & \multicolumn{1}{c}{} & \multicolumn{2}{c}{\pea} & \multicolumn{1}{c}{} & \multicolumn{2}{c}{\aqu} & \multicolumn{1}{c}{} & \multicolumn{2}{c}{\aasc} & \multicolumn{1}{c}{} & \multicolumn{2}{c}{\wil} \\
				\cline{3-4}\cline{6-7}\cline{9-10}\cline{12-13}\cline{15-16}    Scenarios &       & n=15  & n=50  &       & n=15  & n=50  &       & n=15  & n=50  &       & n=15  & n=50  &       & n=15  & n=50 \\
				\hline
				& Mean  & -0.09 & -0.10 &       & 0.00  & 0.00  &       & 0.02  & 0.00  &       & -0.09 & -0.10 &       & -0.08 & -0.09 \\
				& Variance    & 1.02  & 1.00  &       & 1.00  & 1.00  &       & 1.03  & 1.01  &       & 1.01  & 0.99  &       & 1.01  & 1.00 \\
				I-b     & Skewness & -0.12 & -0.06 &       & 0.30  & 0.48  &       & -0.12 & -0.06 &       & -0.11 & -0.05 &       & -0.05 & -0.03 \\
				& Kurtosis & -0.37 & -0.12 &       & -0.34 & 0.24  &       & -0.36 & -0.12 &       & -0.39 & -0.16 &       & -0.39 & -0.12 \\
				& AD Statistics & 22.73 & 25.25 &       & 15.89 & 18.94 &       & 7.00  & 1.68  &       & 22.51 & 25.06 &       & 18.83 & 23.80 \\
				\hline

				& Mean  & -0.05 & -0.05 &       & 0.00  & 0.00  &       & 0.01  & 0.00  &       & -0.05 & -0.05 &       & -0.04 & -0.05 \\
				& Variance    & 1.00  & 1.00  &       & 1.00  & 1.00  &       & 1.01  & 1.00  &       & 1.00  & 1.00  &       & 1.00  & 1.00 \\
				II-b    & Skewness & -0.06 & -0.02 &       & 0.15  & 0.25  &       & -0.06 & -0.02 &       & -0.06 & -0.02 &       & -0.02 & -0.01 \\
				& Kurtosis & -0.41 & -0.12 &       & -0.41 & -0.02 &       & -0.41 & -0.12 &       & -0.42 & -0.13 &       & -0.42 & -0.12 \\
				& AD Statistics & 8.39  & 7.28  &       & 6.87  & 6.03  &       & 4.33  & 1.22  &       & 8.37  & 7.27  &       & 7.41  & 6.93 \\
				\hline
				& Mean  & -0.14 & -0.14 &       & 0.00  & 0.00  &       & 0.02  & 0.01  &       & -0.13 & -0.14 &       & -0.11 & -0.13 \\
				& Variance    & 1.04  & 1.00  &       & 1.00  & 1.00  &       & 1.06  & 1.02  &       & 1.02  & 0.98  &       & 1.04  & 1.00 \\
				III-b    & Skewness & -0.19 & -0.07 &       & 0.42  & 0.72  &       & -0.19 & -0.07 &       & -0.17 & -0.05 &       & -0.08 & -0.03 \\
				& Kurtosis & -0.31 & -0.11 &       & -0.26 & 0.63  &       & -0.30 & -0.10 &       & -0.37 & -0.18 &       & -0.36 & -0.10 \\
				& AD Statistics & 46.04 & 52.09 &       & 28.56 & 40.32 &       & 11.18 & 1.79  &       & 45.09 & 51.41 &       & 37.10 & 49.06 \\
				\hline
					& Mean  & -0.05 & -0.05 &       & 0.00  & 0.00  &       & 0.02  & 0.01  &       & -0.09 & -0.10 &       & -0.08 & -0.10 \\
					& Variance    & 1.00  & 1.00  &       & 1.00  & 1.00  &       & 1.03  & 1.01  &       & 1.01  & 0.99  &       & 1.02  & 1.00 \\
					IV-b     & Skewness & -0.06 & -0.02 &       & 0.30  & 0.53  &       & -0.13 & -0.06 &       & -0.12 & -0.05 &       & -0.05 & -0.02 \\
					& Kurtosis & -0.41 & -0.12 &       & -0.33 & 0.28  &       & -0.36 & -0.09 &       & -0.39 & -0.14 &       & -0.39 & -0.10 \\
					& AD Statistics & 8.39  & 7.28  &       & 14.86 & 21.07 &       & 6.81  & 1.46  &       & 22.13 & 27.13 &       & 18.50 & 25.66 \\
				\hline
				& Mean  & -0.01 & -0.09 &       & 0.00  & 0.00  &       & 0.00  & 0.00  &       & -0.01 & -0.09 &       & -0.01 & -0.08 \\
				& Variance    & 1.00  & 1.00  &       & 1.00  & 1.00  &       & 1.00  & 1.01  &       & 1.00  & 0.99  &       & 1.00  & 1.00 \\
				V-b    & Skewness & 0.00  & -0.05 &       & 0.02  & 0.45  &       & 0.00  & -0.05 &       & 0.00  & -0.05 &       & 0.00  & -0.02 \\
				& Kurtosis & -0.44 & -0.10 &       & -0.44 & 0.19  &       & -0.44 & -0.10 &       & -0.44 & -0.13 &       & -0.44 & -0.10 \\
				& AD Statistics & 3.96  & 21.36 &       & 3.95  & 16.09 &       & 3.87  & 1.52  &       & 3.96  & 21.20 &       & 3.95  & 19.84 \\
				\hline
				& Mean  & -0.08 & -0.09 &       & 0.00  & 0.00  &       & 0.02  & 0.00  &       & -0.08 & -0.09 &       & -0.07 & -0.09 \\
				& Variance    & 1.02  & 1.00  &       & 1.00  & 1.00  &       & 1.03  & 1.01  &       & 1.01  & 0.99  &       & 1.01  & 1.00 \\
				VI-b    & Skewness & -0.12 & -0.05 &       & 0.26  & 0.47  &       & -0.12 & -0.05 &       & -0.11 & -0.05 &       & -0.04 & -0.02 \\
				& Kurtosis & -0.39 & -0.11 &       & -0.38 & 0.21  &       & -0.38 & -0.10 &       & -0.41 & -0.14 &       & -0.41 & -0.10 \\
				& AD Statistics & 19.25 & 22.54 &       & 11.62 & 17.13 &       & 9.78  & 1.55  &       & 19.04 & 22.38 &       & 14.66 & 21.04 \\
				\hline
				& Mean  & -0.05 & -0.05 &       & -0.01 & 0.00  &       & 0.00  & 0.00  &       & -0.05 & -0.05 &       & -0.05 & -0.05 \\
				& Variance    & 1.00  & 1.00  &       & 0.99  & 1.00  &       & 1.00  & 1.00  &       & 1.00  & 1.00  &       & 1.00  & 1.00 \\
				VII-b     & Skewness & -0.04 & -0.03 &       & 0.15  & 0.26  &       & -0.04 & -0.03 &       & -0.04 & -0.03 &       & -0.01 & -0.01 \\
				& Kurtosis & -0.41 & -0.11 &       & -0.41 & -0.02 &       & -0.41 & -0.10 &       & -0.42 & -0.11 &       & -0.42 & -0.11 \\
				& AD Statistics & 8.57  & 6.98  &       & 6.44  & 5.56  &       & 2.52  & 1.26  &       & 8.56  & 6.97  &       & 7.73  & 6.42 \\
				\hline
			\end{tabular}%
		}
		\label{tab:ig}%
	\end{table}%

\begin{landscape}
    \begin{table}
    	\centering
    	\caption{Comparison of the Anderson-Darling statistic for n = 15 and n = 50 - inverse Gaussian regression model.}
    	% Table generated by Excel2LaTeX from sheet 'Planilha2'
    	\scalebox{.6}{
        		\begin{tabular}{ccccrrrrrrr|rrrrrrrrrrr}
    			\hline
    			&       &       &       &       & \multicolumn{1}{c}{\textbf{Standard}} &       & \multicolumn{3}{c}{\textbf{Quantiles}} &       &       &       &       &       &       & \multicolumn{1}{c}{\textbf{Standard}} &       & \multicolumn{3}{c}{\textbf{Quantiles}} &  \\
    			\cline{8-10}\cline{19-21}    \textbf{Scenario} & \textbf{sigma} & \textbf{n} & \textbf{Resíduos} & \multicolumn{1}{c}{\textbf{Mean}} & \multicolumn{1}{c}{\textbf{Deviantion}} & \multicolumn{1}{c}{\textbf{Minimum}} & \multicolumn{1}{c}{\textbf{Q1}} & \multicolumn{1}{c}{\textbf{Q2}} & \multicolumn{1}{c}{\textbf{Q3}} & \multicolumn{1}{c|}{\textbf{Maximum}} & \multicolumn{1}{c}{\textbf{Scenario}} & \multicolumn{1}{c}{\textbf{sigma}} & \multicolumn{1}{c}{\textbf{n}} & \multicolumn{1}{c}{\textbf{Resíduos}} & \multicolumn{1}{c}{\textbf{Mean}} & \multicolumn{1}{c}{\textbf{Deviantion}} & \multicolumn{1}{c}{\textbf{Minimum}} & \multicolumn{1}{c}{\textbf{Q1}} & \multicolumn{1}{c}{\textbf{Q2}} & \multicolumn{1}{c}{\textbf{Q3}} & \multicolumn{1}{c}{\textbf{Maximum}} \\
    		\hline
    		&       &       & \dev  & 22.73 & 11.90 & 4.95  & 17.30 & 18.67 & 27.38 & 48.18 &       &       &       & \multicolumn{1}{c}{\dev} & 3.96  & 1.72  & 1.90  & 2.49  & 4.06  & 4.29  & 8.58 \\
    		&       &       & \pea  & 15.89 & 7.31  & 6.59  & 11.23 & 13.79 & 20.58 & 28.82 &       &       &       & \multicolumn{1}{c}{\pea} & 3.95  & 1.68  & 1.89  & 2.49  & 3.96  & 4.59  & 8.54 \\
    		\textbf{I-b} & \textbf{0.02} & \textbf{15} & \aqu  & 7.00  & 5.41  & 2.53  & 4.18  & 5.32  & 6.48  & 22.80 & \multicolumn{1}{c}{\textbf{V-b}} & \multicolumn{1}{c}{\textbf{0.02}} & \multicolumn{1}{c}{\textbf{15}} & \multicolumn{1}{c}{\aqu} & 3.87  & 1.71  & 2.06  & 2.65  & 3.59  & 4.80  & 8.75 \\
    		&       &       & \aasc & 22.51 & 11.72 & 4.94  & 17.20 & 18.44 & 26.99 & 47.59 &       &       &       & \multicolumn{1}{c}{\aasc} & 3.96  & 1.72  & 1.90  & 2.49  & 4.06  & 4.29  & 8.58 \\
    		&       &       & \wil  & 18.83 & 10.11 & 4.83  & 14.24 & 16.88 & 21.32 & 40.40 &       &       &       & \multicolumn{1}{c}{\wil} & 3.95  & 1.70  & 1.89  & 2.47  & 4.03  & 4.31  & 8.57 \\
    		\hline
    		&       &       & \dev  & 25.25 & 20.46 & 2.30  & 10.83 & 16.80 & 37.06 & 93.32 &       &       &       & \multicolumn{1}{c}{\dev} & 21.36 & 18.89 & 2.42  & 9.49  & 16.13 & 26.89 & 113.98 \\
    		&       &       & \pea  & 18.94 & 14.74 & 2.80  & 8.35  & 12.81 & 26.71 & 70.49 &       &       &       & \multicolumn{1}{c}{\pea} & 16.09 & 12.47 & 2.04  & 8.66  & 11.99 & 18.82 & 72.26 \\
    		\textbf{I-b} & \textbf{0.02} & \textbf{50} & \aqu  & 1.68  & 1.30  & 0.38  & 0.78  & 1.33  & 1.98  & 5.06  & \multicolumn{1}{c}{\textbf{V-b}} & \multicolumn{1}{c}{\textbf{0.02}} & \multicolumn{1}{c}{\textbf{50}} & \multicolumn{1}{c}{\aqu} & 1.52  & 1.33  & 0.34  & 0.70  & 1.06  & 1.79  & 7.16 \\
    		&       &       & \aasc & 25.06 & 20.27 & 2.31  & 10.82 & 16.82 & 36.81 & 92.65 &       &       &       & \multicolumn{1}{c}{\aasc} & 21.20 & 18.57 & 2.40  & 9.49  & 16.06 & 26.80 & 111.49 \\
    		&       &       & \wil  & 23.80 & 19.76 & 1.85  & 9.54  & 16.29 & 34.85 & 89.89 &       &       &       & \multicolumn{1}{c}{\wil} & 19.84 & 16.63 & 1.89  & 8.98  & 14.28 & 25.90 & 94.32 \\
    		\hline
    		&       &       & \dev  & 22.37 & 14.51 & 6.31  & 12.10 & 20.37 & 26.50 & 55.21 &       &       &       & \multicolumn{1}{c}{\dev} & 19.25 & 12.94 & 4.03  & 12.02 & 13.65 & 23.28 & 50.91 \\
    		&       &       & \pea  & 14.86 & 9.26  & 2.16  & 9.38  & 13.79 & 20.05 & 32.75 &       &       &       & \multicolumn{1}{c}{\pea} & 11.62 & 4.47  & 3.00  & 10.01 & 10.36 & 14.66 & 19.58 \\
    		\textbf{II-b} & \textbf{0.5} & \textbf{15} & \aqu  & 6.81  & 2.70  & 3.83  & 4.89  & 5.63  & 8.60  & 11.95 & \multicolumn{1}{c}{\textbf{VI-b}} & \multicolumn{1}{c}{\textbf{0.02}} & \multicolumn{1}{c}{\textbf{15}} & \multicolumn{1}{c}{\aqu} & 9.78  & 13.55 & 3.02  & 4.40  & 6.36  & 8.02  & 57.70 \\
    		&       &       & \aasc & 22.13 & 14.27 & 6.28  & 12.04 & 20.12 & 26.36 & 54.41 &       &       &       & \multicolumn{1}{c}{\aasc} & 19.04 & 12.63 & 4.02  & 11.93 & 13.56 & 23.12 & 49.58 \\
    		&       &       & \wil  & 18.50 & 13.00 & 3.52  & 9.79  & 15.47 & 22.53 & 47.69 &       &       &       & \multicolumn{1}{c}{\wil} & 14.66 & 7.67  & 2.29  & 10.06 & 12.04 & 20.77 & 30.29 \\
    		\hline
    		&       &       & \dev  & 27.32 & 16.62 & 6.20  & 14.51 & 21.95 & 39.21 & 69.40 &       &       &       & \multicolumn{1}{c}{\dev} & 22.54 & 16.70 & 3.30  & 12.09 & 18.85 & 28.34 & 79.51 \\
    		&       &       & \pea  & 21.07 & 12.16 & 5.16  & 11.45 & 16.96 & 31.67 & 52.55 &       &       &       & \multicolumn{1}{c}{\pea} & 17.13 & 12.07 & 2.14  & 9.72  & 13.20 & 21.80 & 59.69 \\
    		\textbf{II-b} & \textbf{0.5} & \textbf{50} & \aqu  & 1.46  & 0.96  & 0.36  & 0.77  & 1.14  & 1.86  & 5.04  & \multicolumn{1}{c}{\textbf{VI-b}} & \multicolumn{1}{c}{\textbf{0.02}} & \multicolumn{1}{c}{\textbf{50}} & \multicolumn{1}{c}{\aqu} & 1.55  & 1.21  & 0.39  & 0.72  & 1.02  & 2.08  & 6.30 \\
    		&       &       & \aasc & 27.13 & 16.46 & 6.21  & 14.49 & 21.79 & 38.98 & 68.90 &       &       &       & \multicolumn{1}{c}{\aasc} & 22.38 & 16.49 & 3.27  & 12.02 & 18.76 & 28.19 & 78.25 \\
    		&       &       & \wil  & 25.66 & 15.85 & 5.19  & 13.72 & 20.28 & 36.79 & 65.91 &       &       &       & \multicolumn{1}{c}{\wil} & 21.04 & 15.36 & 2.40  & 11.49 & 18.15 & 27.25 & 73.74 \\
    		\hline
    		&       &       & \dev  & 8.39  & 4.01  & 2.26  & 5.62  & 8.53  & 10.65 & 15.39 &       &       &       & \multicolumn{1}{c}{\dev} & 8.57  & 3.94  & 3.65  & 5.41  & 7.03  & 8.86  & 19.46 \\
    		&       &       & \pea  & 6.87  & 2.59  & 2.75  & 5.16  & 6.50  & 8.83  & 11.10 &       &       &       & \multicolumn{1}{c}{\pea} & 6.44  & 2.84  & 3.43  & 4.59  & 5.71  & 6.22  & 15.10 \\
    		\textbf{III-b} & \textbf{0.01} & \textbf{15} & \aqu  & 4.33  & 2.11  & 2.02  & 2.83  & 3.70  & 5.42  & 10.13 & \multicolumn{1}{c}{\textbf{VII-b}} & \multicolumn{1}{c}{\textbf{0.02}} & \multicolumn{1}{c}{\textbf{15}} & \multicolumn{1}{c}{\aqu} & 2.52  & 2.32  & 1.81  & 3.19  & 4.62  & 5.77  & 9.66 \\
    		&       &       & \aasc & 8.37  & 4.00  & 2.25  & 5.60  & 8.52  & 10.57 & 15.37 &       &       &       & \multicolumn{1}{c}{\aasc} & 8.56  & 3.94  & 3.63  & 5.38  & 7.03  & 8.84  & 19.44 \\
    		&       &       & \wil  & 7.41  & 3.57  & 2.18  & 4.71  & 7.80  & 9.61  & 13.49 &       &       &       & \multicolumn{1}{c}{\wil} & 7.73  & 3.67  & 3.34  & 4.34  & 5.84  & 7.21  & 18.40 \\
    		\hline
    		&       &       & \dev  & 7.28  & 5.83  & 0.38  & 3.24  & 5.56  & 9.18  & 26.78 &       &       &       & \multicolumn{1}{c}{\dev} & 6.98  & 5.28  & 1.83  & 4.16  & 5.43  & 7.51  & 33.83 \\
    		&       &       & \pea  & 6.03  & 4.32  & 1.20  & 2.66  & 4.29  & 8.81  & 19.87 &       &       &       & \multicolumn{1}{c}{\pea} & 5.56  & 2.69  & 2.55  & 4.19  & 4.90  & 5.80  & 18.06 \\
    		\textbf{III-b} & \textbf{0.01} & \textbf{50} & \aqu  & 1.22  & 0.91  & 0.21  & 0.57  & 0.84  & 1.62  & 3.83  & \multicolumn{1}{c}{\textbf{VII-b}} & \multicolumn{1}{c}{\textbf{0.02}} & \multicolumn{1}{c}{\textbf{50}} & \multicolumn{1}{c}{\aqu} & 1.26  & 0.79  & 0.35  & 0.72  & 0.99  & 1.59  & 3.72 \\
    		&       &       & \aasc & 7.27  & 5.81  & 0.39  & 3.24  & 5.54  & 9.14  & 26.79 &       &       &       & \multicolumn{1}{c}{\aasc} & 6.97  & 5.26  & 1.81  & 4.14  & 5.44  & 7.50  & 33.68 \\
    		&       &       & \wil  & 6.93  & 5.65  & 0.38  & 3.09  & 5.24  & 8.69  & 26.06 &       &       &       & \multicolumn{1}{c}{\wil} & 6.42  & 4.26  & 1.69  & 3.86  & 5.03  & 7.23  & 23.83 \\
    		\hline
    		&       &       & \dev  & 46.04 & 25.83 & 11.37 & 27.46 & 44.21 & 65.43 & 89.61 &       &       &       &       &       &       &       &       &       &       &  \\
    		&       &       & \pea  & 28.56 & 16.25 & 5.94  & 17.62 & 26.13 & 41.13 & 53.18 &       &       &       &       &       &       &       &       &       &       &  \\
    		\textbf{IV-b} & \textbf{0.03} & \textbf{15} & \aqu  & 11.18 & 3.96  & 6.02  & 8.30  & 11.53 & 13.60 & 19.44 &       &       &       &       &       &       &       &       &       &       &  \\
    		&       &       & \aasc & 45.09 & 25.05 & 11.11 & 27.12 & 43.53 & 63.89 & 87.26 &       &       &       &       &       &       &       &       &       &       &  \\
    		&       &       & \wil  & 37.10 & 22.63 & 9.23  & 19.92 & 38.44 & 51.68 & 75.13 &       &       &       &       &       &       &       &       &       &       &  \\
    		\cline{1-11}      &       &       & \dev  & 52.09 & 33.93 & 11.63 & 24.63 & 39.33 & 75.39 & 125.73 &       &       &       &       &       &       &       &       &       &       &  \\
    		&       &       & \pea  & 40.32 & 26.26 & 5.45  & 20.48 & 31.17 & 56.34 & 106.07 &       &       &       &       &       &       &       &       &       &       &  \\
    		\textbf{IV-b} & \textbf{0.03} & \textbf{50} & \aqu  & 1.79  & 1.03  & 0.35  & 1.02  & 1.59  & 2.32  & 4.49  &       &       &       &       &       &       &       &       &       &       &  \\
    		&       &       & \aasc & 51.41 & 33.27 & 11.60 & 24.49 & 39.11 & 74.52 & 123.86 &       &       &       &       &       &       &       &       &       &       &  \\
    		&       &       & \wil  & 49.06 & 32.52 & 8.40  & 23.27 & 37.83 & 71.68 & 119.60 &       &       &       &       &       &       &       &       &       &       &  \\
    		\cline{1-11}\   \end{tabular}%
    		\label{adig}%
    }	
    	\end{table}%
    
\end{landscape}

\section{Application}\label{apps}

In Section \ref{simu}, we studied whether the distribution of the residuals considered in this paper are well approximated by the standard normal distribution. 
Other essential properties for residuals are their ability to identify model misspecification and outliers.
We used two real datasets to study the residuals considered in this paper according to these aspects.
In this section, we compare the \aqu\, with the two most commonly used residual in the GLMs, the  \dev\, and the \pea\,.
Applications were performed using the \textit{gamlss} package and the \textit{glm} function of the R software.

\subsection{Oil dataset}

\begin{figure}[H]
	\centering
	\includegraphics[width=0.35\linewidth]{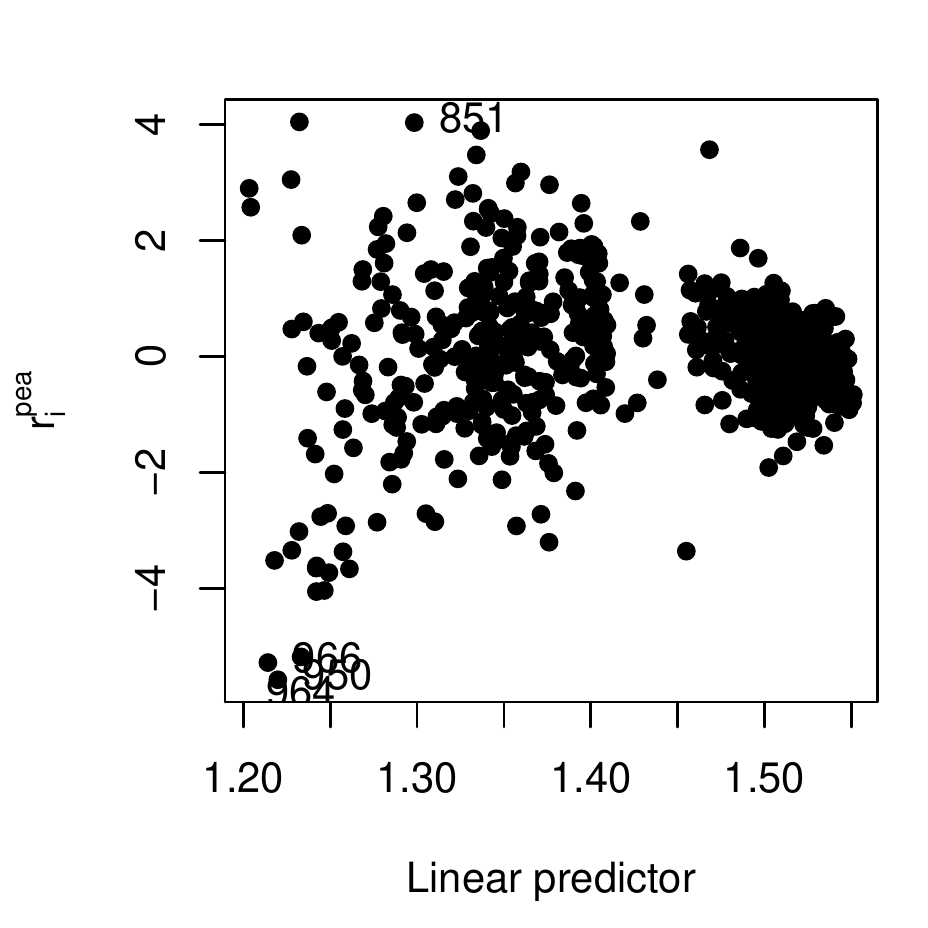}
	\includegraphics[width=0.35\linewidth]{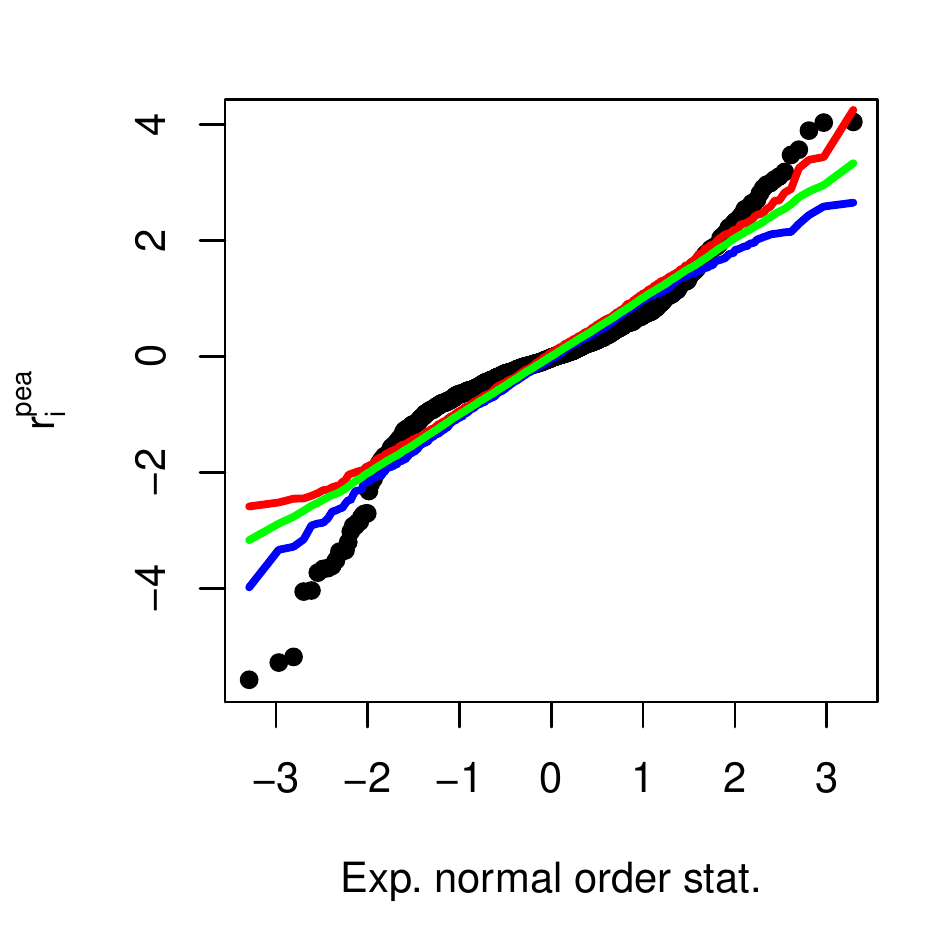}\\
	\includegraphics[width=0.35\linewidth]{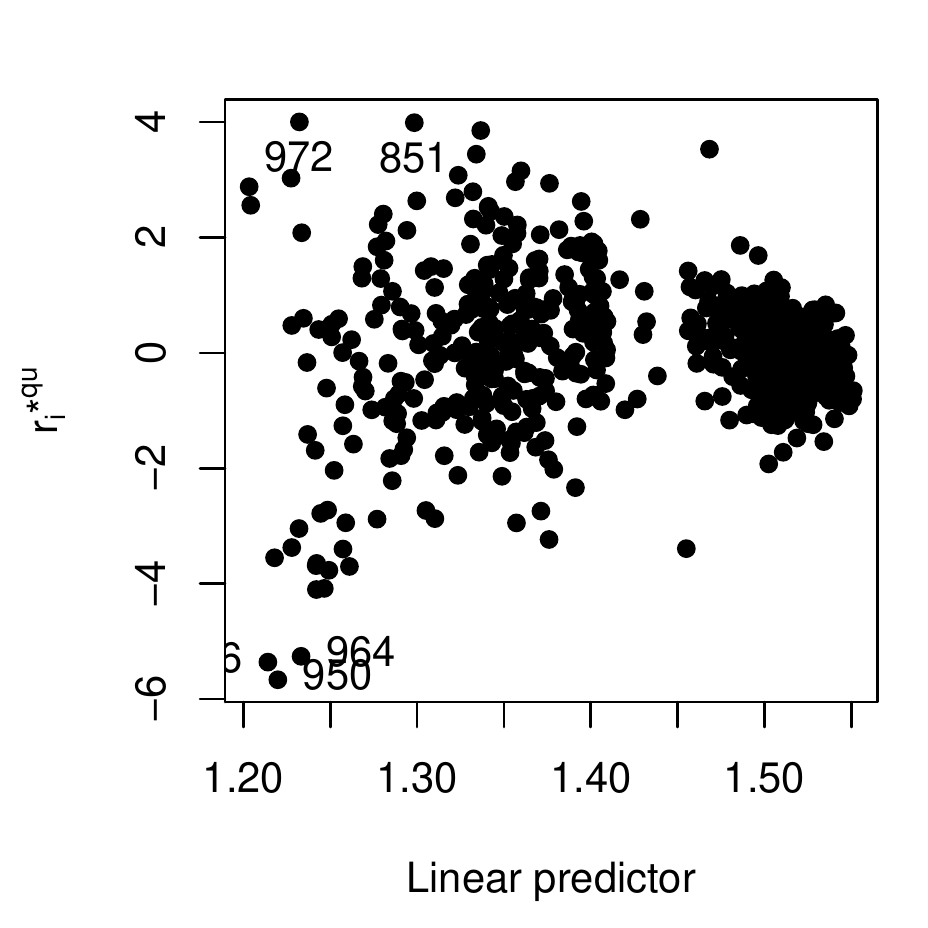}
	\includegraphics[width=0.35\linewidth]{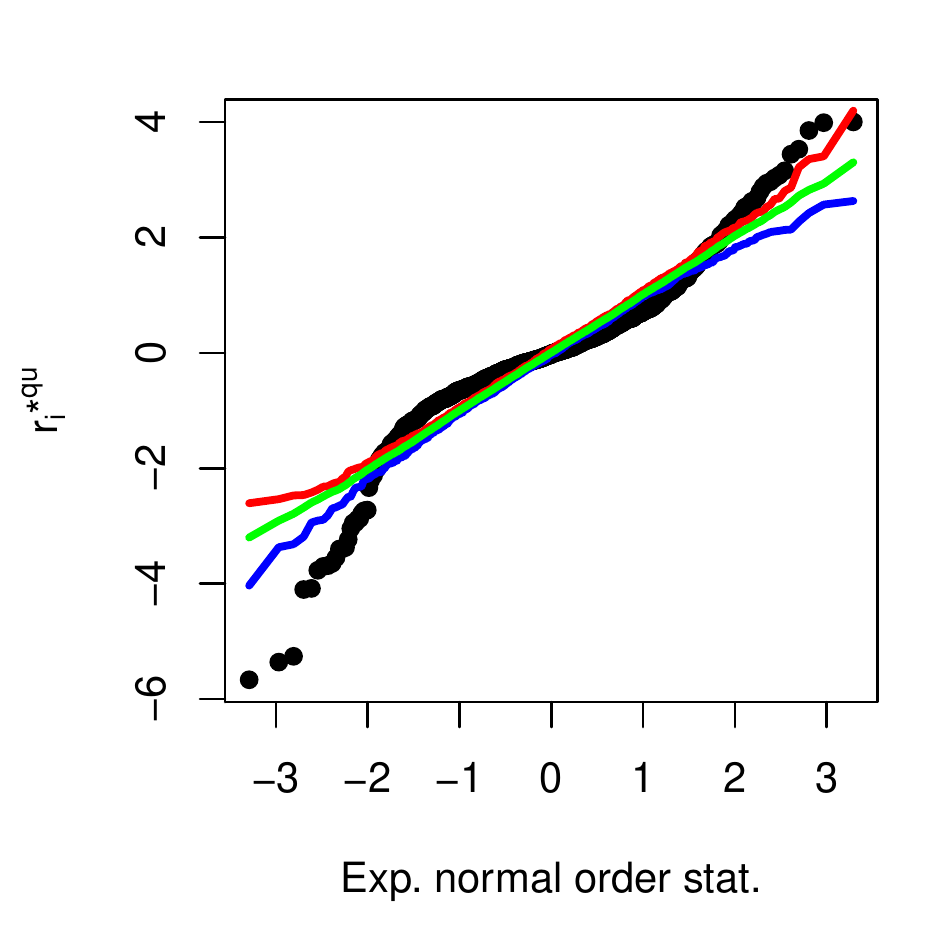}\\
	\caption{Residuals against linear predictor (left) and normal residual plots with simulated envelope (right) for oil dataset.}
	\label{segreg}
\end{figure}

The first application uses a database with 1000 observations about the daily prices of WTI (West Texas Intermediate) oil price traded by NYMEX(New York Mercantile Exchange). We used this dataset to investigate whether the \dev, \pea and \aqu \, are able to identify model misspecification.
The response variable is the price of oil. Here, we use two explanatory variables: the  lagged version of the response variable and the log price of front month heating oil contract traded by NYMEX, as proposed by \cite{rigby2015}. 
We fitted inverse Gaussian regression model with logit link function for the oil data and obtained the three sets of residuals.
\cite{rigby2015} proposed the Sinh-Arcsinh (SHASH) distribution and splines to obtain a reasonable fit for this response variable. Therefore, lack of fit is expected when we fit the inverse Gaussian regression model. 
Figure \ref{segreg} presents a plot of residuals versus linear predictor and a normal residual plot with simulated envelope \citep{atkinson} for \dev, \pea and \aqu.
Clearly, the plots for the three residuals suggest lack of fit for the inverse Gaussian regression model, indicating that these residuals are able to identify this kind of problem. A similar analysis was performed using the gamma regression model. The conclusions were the same and the plots were omitted for the sake of brevity.
\begin{figure}[H]
	\nonumber
	\centering
	\includegraphics[width=0.35\linewidth]{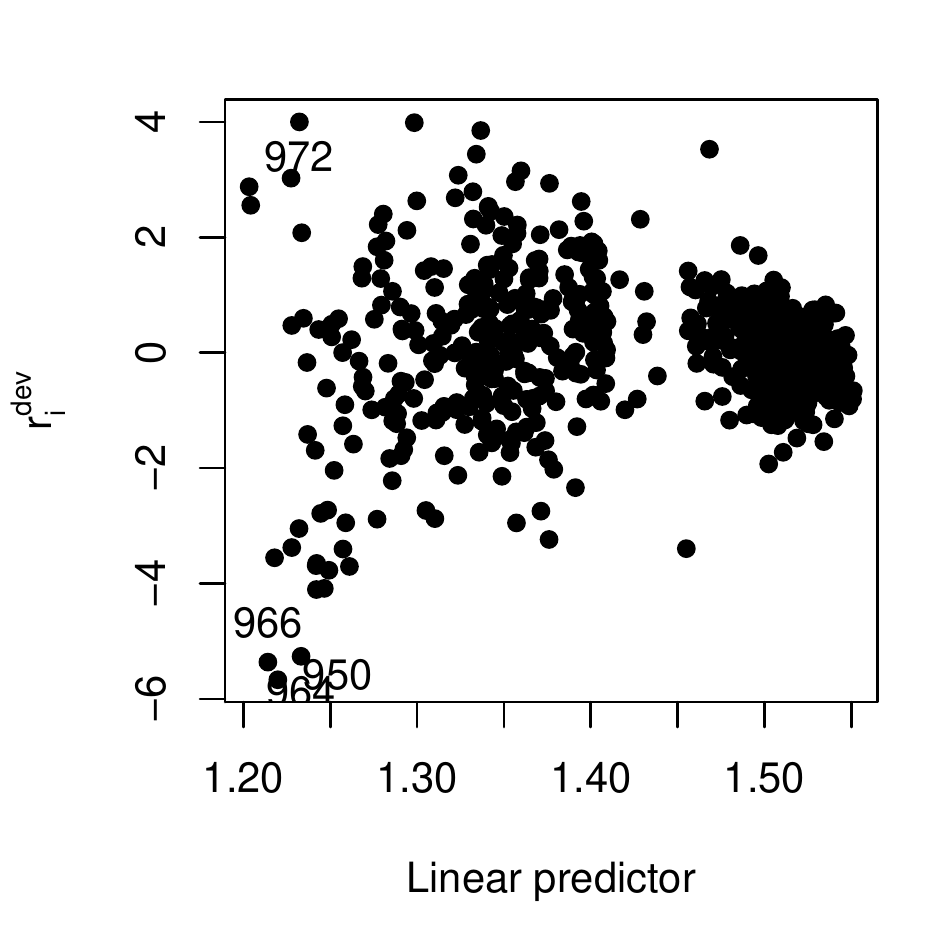}
	\includegraphics[width=0.35\linewidth]{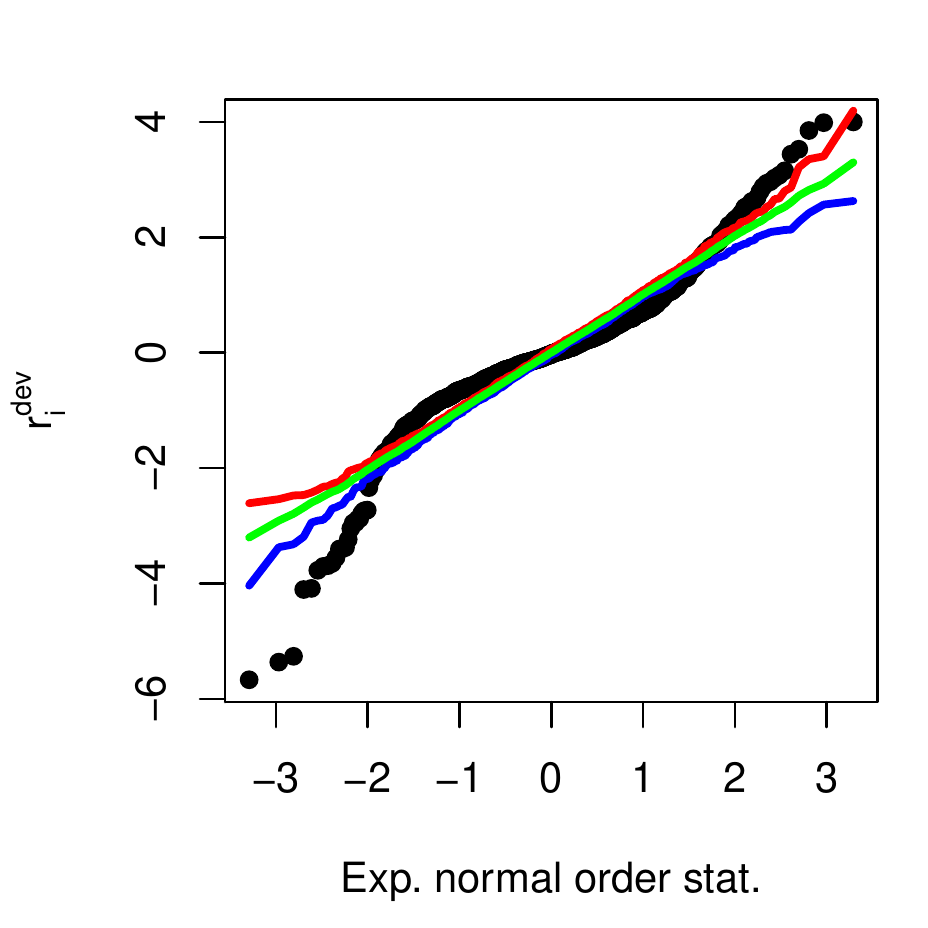}\\
\end{figure}

\subsection{Turbine dataset}
\begin{figure}[H]
	\centering
	\includegraphics[width=0.35\linewidth]{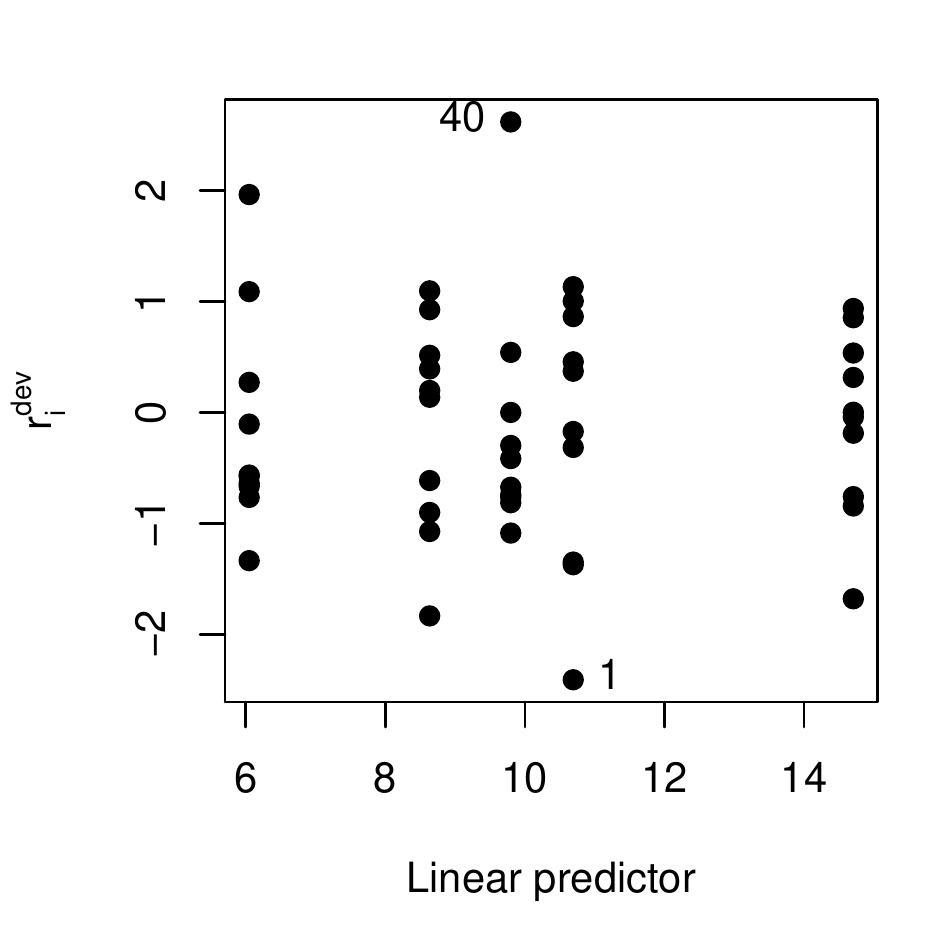}
	\includegraphics[width=0.35\linewidth]{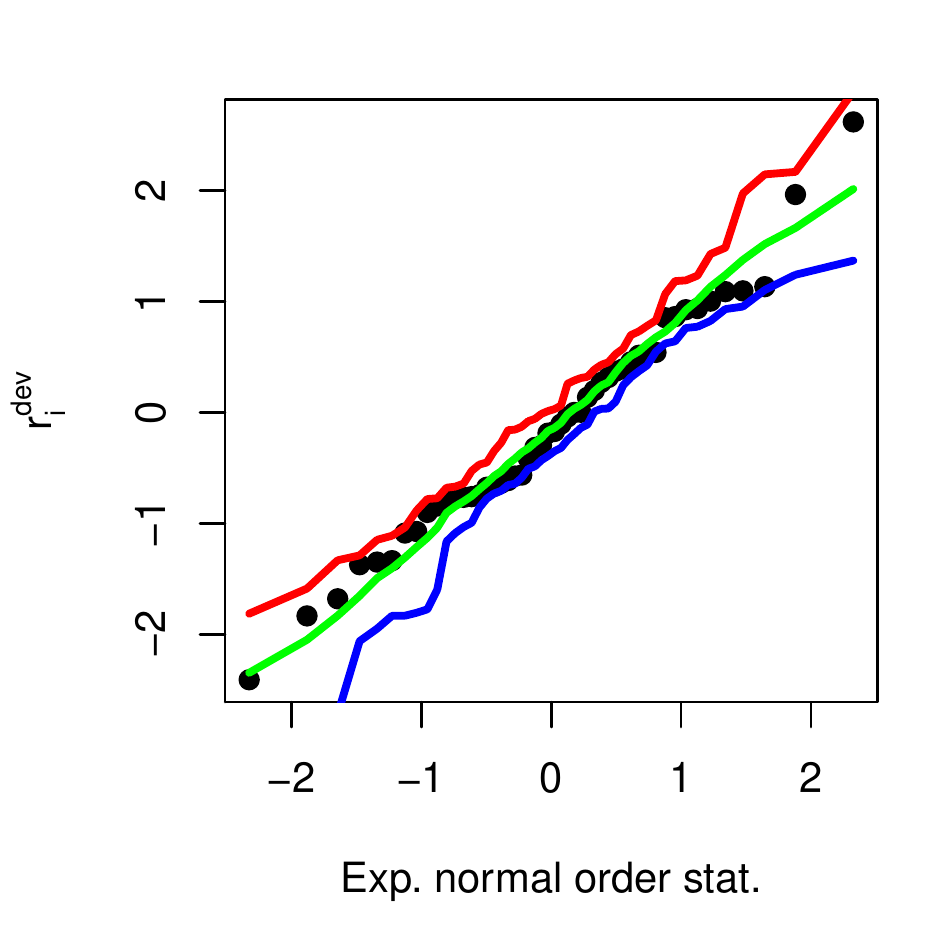}\\
	\includegraphics[width=0.35\linewidth]{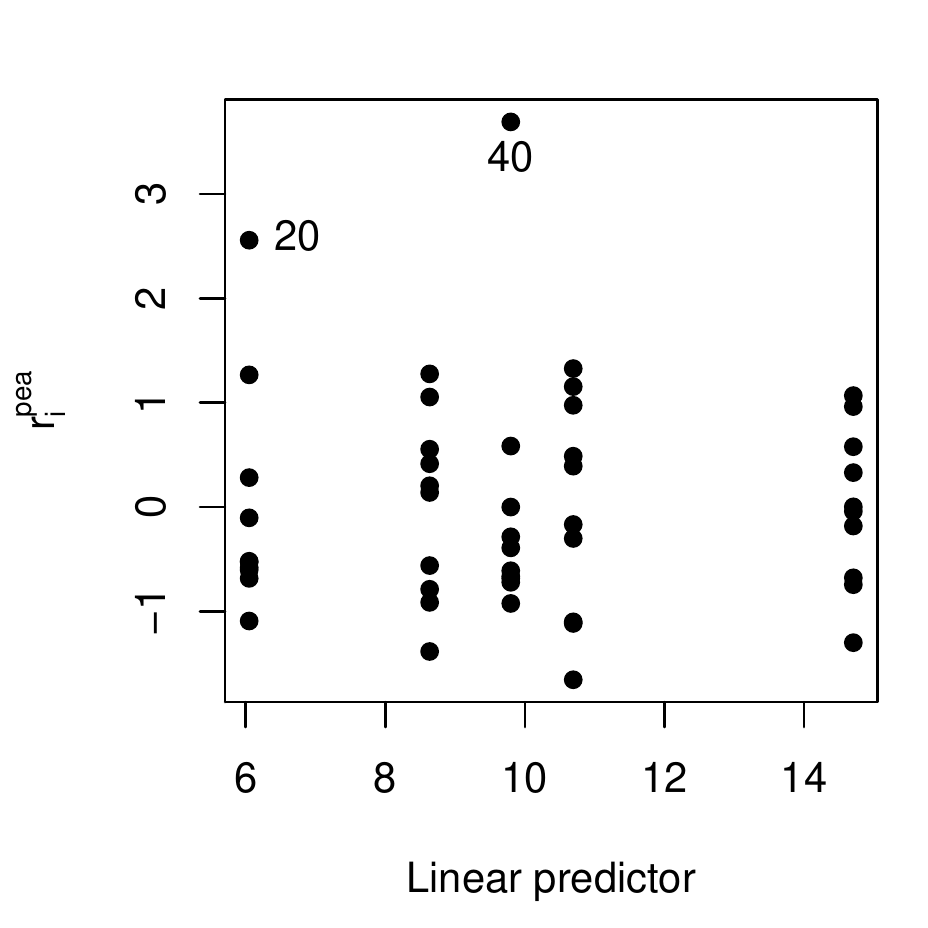}
	\includegraphics[width=0.35\linewidth]{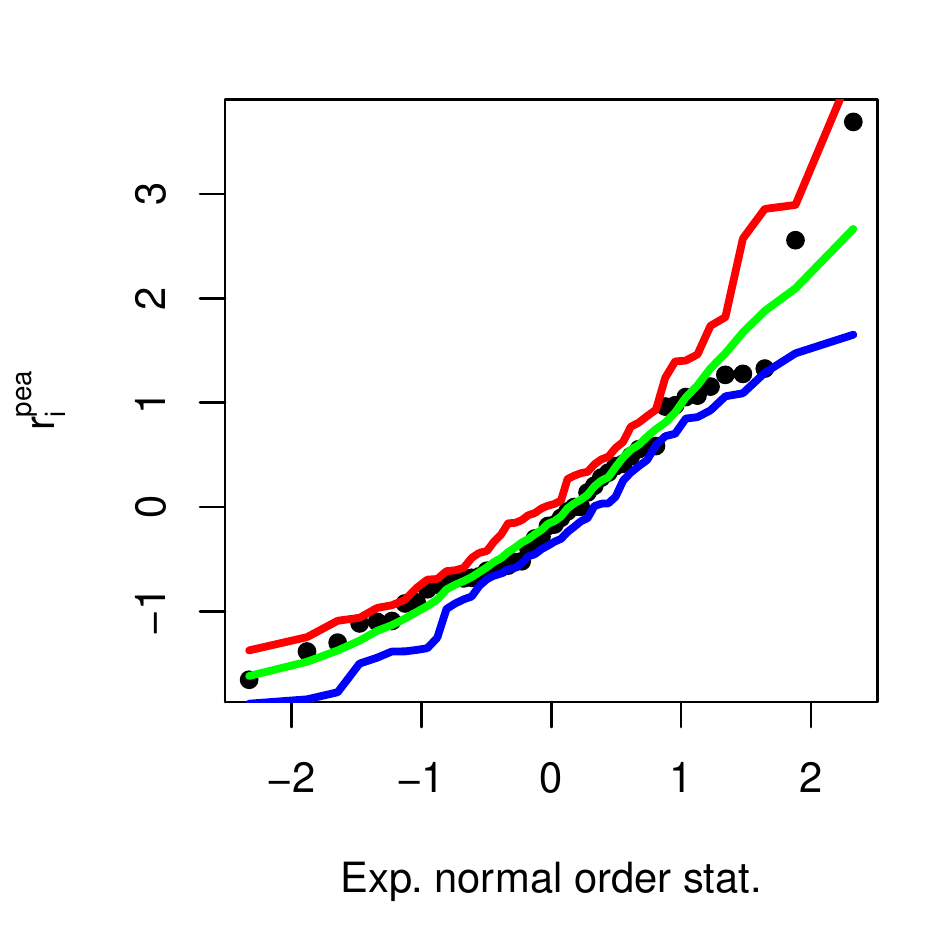}\\
	\includegraphics[width=0.35\linewidth]{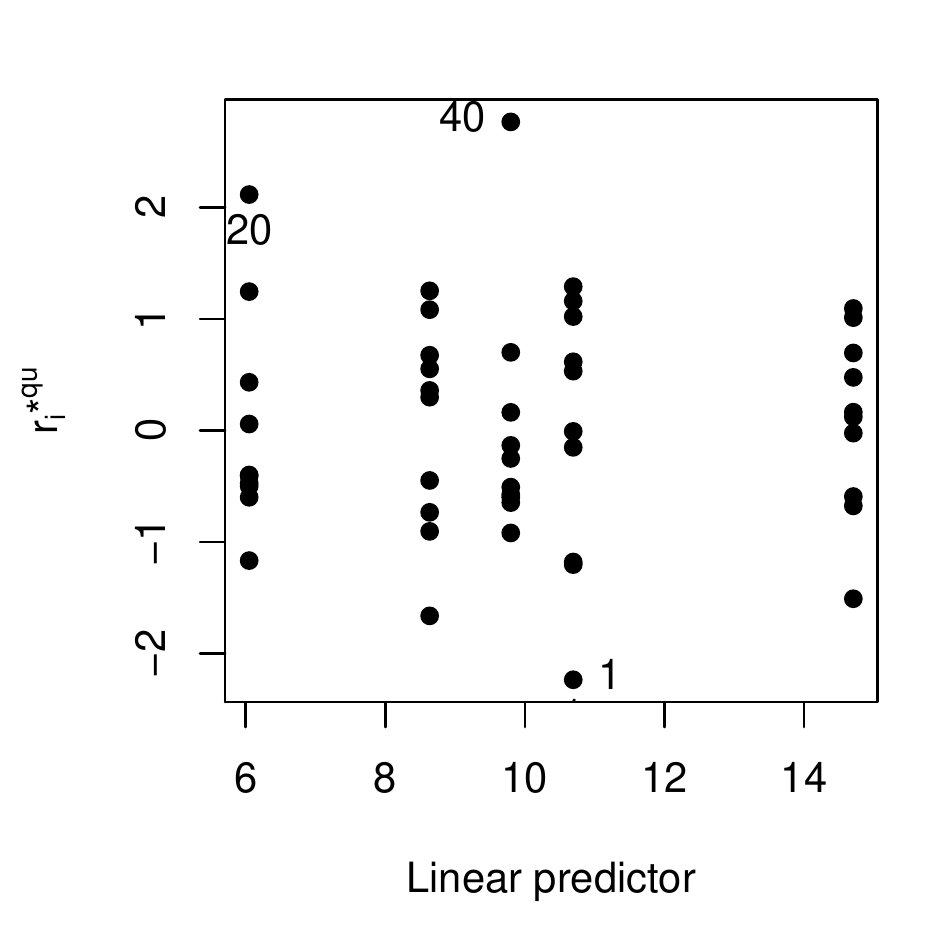}
	\includegraphics[width=0.35\linewidth]{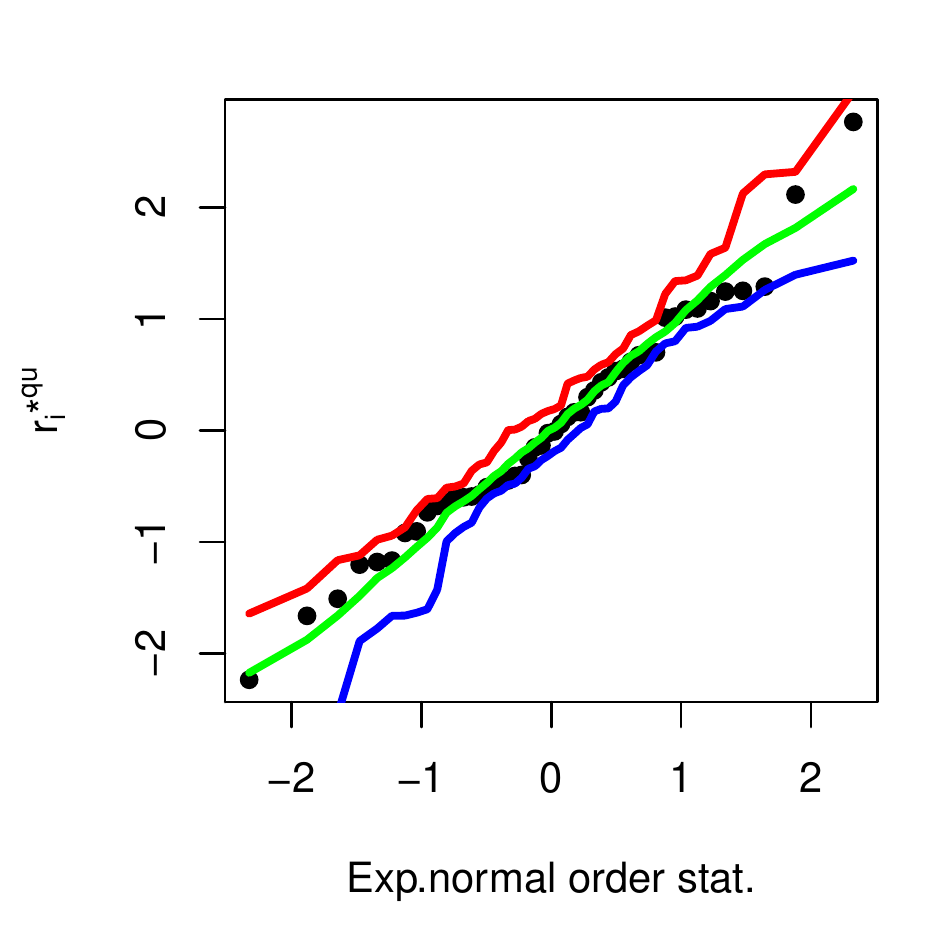}
	\caption{Residuals against linear predictor (left) and half-normal residual plots with simulated envelope (right) for turbine data.}
	\label{turbine}
\end{figure}

The second real dataset consists of the time to evaluate the performance of five types of high-speed turbines for plane engines \citep{lawless}. 
The response variable is the time (in million of cycles units) until the loss of velocity and the covariate is the type of turbine.
Our focus with this application is to check if \dev, \pea, \aqu\, are good residuals to perform outlier identification.
We fitted gamma regression model with identity link function to this dataset and obtained the three set of residuals.
Figure \ref{turbine} presents a plot of residuals against linear predictors and a normal residual plot with simulated envelope for \dev, \pea and \aqu.
None of the residuals suggest model misspecification.

For the three residuals, observation $40$ has the highest value and observation $1$ has the smallest. However, the value of the residual for observation $40$ is considerably greater for \pea\, than for the other residuals and the absolute value of the residual for observation $1$ is substantially smaller for \pea\, than for \dev\, and \aqu\,. In the simulation studies, we noted that the distribution of \pea\ is right skewed and this explains the mentioned result. As a consequence, \pea\ may incorrectly identify as an outlier an observation with positive value for the residual and may not identify an outlier with value of \pea\ negative. On the other hand, the values of \dev\, and \aqu\, are close for all observations. However, the values of the positive residual are slightly greater for \aqu\ than for \dev\, and the absolute values of the negative residuals are slightly smaller for \aqu\ than for \dev\,. As \dev\, has mean far from zero than \aqu\, and similar variance, skewness and excess kurtosis, the latter seems to be a better measure of the discrepancy of an observation and, consequently, better to perform outlier identification.

\section{Conclusions}\label{conc}

In this work, we introduced the adjusted quantile residual to perform diagnostic analysis in generalized linear models. We compared the distribution of the adjusted quantile residual with four other residuals using Monte Carlos simulation studies. Additionally, we used two applications to investigate if the residuals are able to identify model misspecification and outliers.

It is very desirable to find a residual whose distribution is well approximated by the standard normal distribution.
Our simulation studies suggest that the adjusted quantile residual's distribution better approximates to the standard normal distribution in all scenarios than the other residual's distribution, specially when sample size is not small. Simulation results also suggest that, when variance increases, all residuals worsen regarding normal approximation. However, when variance is high, the distribution of the quantile residual is much better
approximated by the standard normal distribution than that of the other residuals, even in small samples. 

The applications investigated other properties of the residuals.
The first application suggested that the three residuals considered in the analysis can detect lack of fit in generalized linear models.
Based on the second application, the adjusted quantile residual seems to be the best to perform outlier identification.

The standardized Pearson residual and the standardized deviance residual are calculated by many packages and statistical softwares in their generalized linear models routine, but commonly the quantile residual is not implemented. However, adjusted quantile residual is simple and easy to calculate using any statistical package. Considering the results of the simulations studies, the applications and its simplicity, the adjusted quantile residual is a better choice of to perform diagnostic analysis in generalized linear models than the competing residuals.

\section{Acknowledgements}
The authors thank "Coordenação de Aperfeiçoamento de Pessoal de Nível Superior" (CAPES) for the financial support received for this project.

\section{References}

%\singlespacing

\bibliographystyle{agsm}
\bibliography{bibliografia}

\end{document}